\documentclass[useAMS,usenatbib]{mnras}
\usepackage{graphicx}
\usepackage{txfonts}
\usepackage{color} 

\title[Calibrations for abundance determinations]
      {New calibrations for abundance determinations in H\,{\sc ii} regions} 
      
\author[Pilyugin \& Grebel]
       {L.~S.~Pilyugin$^{1,2,3}$,
        E.~K.~Grebel$^{2}$ \\
       $^{1}$ Main Astronomical Observatory
             of National Academy of Sciences of Ukraine,
             27 Zabolotnogo str., 03680 Kiev, Ukraine \\   
       $^{2}$ Astronomisches Rechen-Institut, Zentrum f\"{u}r Astronomie 
             der Universit\"{a}t Heidelberg, 
             M\"{o}nchhofstr.\ 12--14, 69120 Heidelberg, Germany \\
       $^{3}$ Kazan Federal University, 18 Kremlyovskaya St., 420008, Kazan. Russian Federation
        }

\date{Accepted 2016 January 26. Received 2016 January 24; in original form 2016 January 05}

\begin{document}

\maketitle

\begin{abstract} 
Simple relations for deriving the oxygen abundance in H\,{\sc ii}
regions with intensities of the three strong emission lines $R_{2}$,
$R_{3}$, and $N_{2}$ ($R$ calibration) or $S_{2}$, $R_{3}$, and
$N_{2}$ ($S$ calibration) in their spectra are suggested.  A sample of
313 reference H\,{\sc ii} regions of the counterpart method ($C$
method) is used as calibrating data points. Relations for the
determination of nitrogen abundances, the $R$ calibration, are also
constructed.  We find that the oxygen and nitrogen abundances in
high-metallicity  H\,{\sc ii} regions can be estimated using the
intensities of the two strong lines $R_{2}$ and $N_{2}$ (or $S_{2}$
and $N_{2}$ for oxygen) only. The corresponding two-dimensional
relations are provided. There are considerable advantages of the
suggested calibration relations as compared to the existing ones.
First, the oxygen and nitrogen abundances estimated through the
suggested calibrations agree with the $T_{e}$-based abundances within
$\sim 0.1$ dex over the whole metallicity range, i.e., the relative 
accuracy of the calibration-based abundances is 0.1 dex. Although we
constructed distinct relations for high- and low-metallicity objects,
the separation between these two can be simply obtained from the
intensity of the $N_{2}$ line. Moreover, the applicability ranges of
the high- and low-metallicity relations overlap for adjacent
metallicities, i.e., the transition zone disappears.  Second, the
oxygen abundances produced by the two suggested calibrations are in
remarkable agreement with each other. In fact, the $R$-based and
$S$-based oxygen abundances agree within $\sim$0.05 dex in the
majority of cases for more than three thousand H\,{\sc ii} region
spectra. 
\end{abstract}


\section{Introduction}

Reliable chemical abundance determinations are essential for a wide
variety of investigations of galaxies and their evolution. 
For example, on global scales they are one of the key parameters in
the study of the luminosity-metallicity and mass-metallicity relations
of galaxies, their evolution with time, and their dependence on
environment or star formation rate \citep[e.g.,][to just name a few of the
many studies]{Lequeux1979, Grebel2003, Tremonti2004, Erb2006,
Panter2008, Zahid2012, Sanchez2013, Peng2014, Zahid2014, Izotov2015}.
Local measurements within galaxies reveal the position-dependent
metallicity of the chosen tracer population and may show abundance
gradients, which in turn hold clues about galaxy evolution
\citep[e.g.,][]{Searle1971, Janes1979, Maciel1999, Harbeck2001,
Andrievsky2002, Chen2003, Mehlert2003, Cioni2009, Haschke2012,
Boeche2013, Boeche2014, Pilyugin2014, Pilyugin2015}.  
  
The metallicity of star-forming galaxies at the present epoch can
be estimated from the emission-line spectra of H\,{\sc ii} regions,
which are easily obtained even over greater distances. It is believed
that the direct $T_{e}$ method \citep[e.g.,][]{Dinerstein1990}
provides the most reliable abundance determinations in H\,{\sc ii}
regions. Abundance determinations through the direct $T_{e}$ method
require high-precision spectroscopy of H\,{\sc ii} regions in order
to detect the weak auroral lines such as [O\,{\sc iii}]$\lambda$4363
or/and [N\,{\sc ii}]$\lambda$5755.  Unfortunately, these auroral lines
are often rather faint and thus may be detected only in the spectra of
a limited number of H\,{\sc ii} regions.  The abundances in other
H\,{\sc ii} regions are then usually estimated through the method
suggested by \citet{Pagel1979} and \citet{Alloin1979}.  The idea of
this method (traditionally called the strong-line method) is to
establish the relation between the (oxygen) abundance in an H\,{\sc
ii} region and some combination of the intensities of strong emission
lines in its spectrum, i.e., the combination of the intensities of
these strong lines is calibrated in terms of the metallicity of the
H\,{\sc ii} region.  Therefore, such a relation is usually called a
``calibration'' and serves to convert metallicity-sensitive
emission-line combinations into metallicity estimations.  

Numerous calibrations based on the emission lines of different
elements were suggested \citep[][among many
others]{Edmunds1984,Dopita1986,McGaugh1991,Zaritsky1994,Pilyugin2000,Pilyugin2001a,Kewley2002,
Pettini2004,Tremonti2004,Pilyugin2005,Stasinska2006,Pilyugin2010,Marino2013,MoralesLuis2014}.
A prominent characteristic of all the calibrations is that there is no
unique relation that is applicable across the whole range of
metallicities of H\,{\sc ii} regions. Instead, a calibration relation
for a limited range of metallicities or distinct calibration relations
for different intervals of metallicities (usually at high or at low
metallicities) are constructed.  One has to know {\em a priori} to
which interval of metallicity the H\,{\sc ii} region belongs in order
to choose the relevant calibration relation 
\citep[e.g.,][]{Kewley2002,Blanc2015}. This can result in a
wrong choice of the calibration relation and, as a consequence, in
large uncertainties in the abundances of the H\,{\sc ii} regions.
This problem is particularly difficult for H\,{\sc ii} regions that
lie near the boundary of the applicability of the calibration
relation. 

The most important characteristic of the calibration is a sample of
calibrating data points to be used in the construction of the
calibration relation.  Grids of photoionization models of H\,{\sc ii}
regions can be used to establish a relation between strong-line
intensities and oxygen abundances
\citep[e.g.,][]{McCall1985,Dopita1986,McGaugh1991,Kewley2002,Dopita2013}.  
Such calibrations are usually referred to as theoretical or model
calibrations.  On the other hand, a sample of H\,{\sc ii} regions in
which the oxygen abundances are determined through the direct $T_{e}$
method can serve as basis of a calibration
\citep[e.g.,][]{Pilyugin2000,Pilyugin2001a,Pilyugin2010,Marino2013}.
Such calibrations are usually called empirical calibrations. There are
also a hybrid calibrations where both the H\,{\sc ii} regions with
directly measured abundances and the photoionization models of H\,{\sc
ii} regions are used \citep[e.g.,][]{Pettini2004}. 

There are large systematic discrepancies between the abundance values
produced by different published calibrations. The theoretical (or
model) calibrations generally produce oxygen abundances that are by
factors of 1.5 to 5 higher than those derived through the direct
$T_{e}$ method or through empirical calibrations
\citep[e.g.,][]{Kennicutt2003,Pilyugin2003a,Kewley2008,Bresolin2009,
Moustakas2010,LopezSanchez2010,LopezSanchez2012}. 
Thus, at the present time there is no absolute scale for metallicities
of H\,{\sc ii} regions. The empirical calibrations have advantages as
compared to the theoretical calibrations. The empirical metallicity
scale is well defined in terms of the abundances in H\,{\sc ii}
regions derived through the direct $T_e$ method, i.e., in that sense
the empirical metallicity scale is absolute. The abundances in H\,{\sc
ii} regions obtained through the different empirical calibrations are
compatible with each other as well as with the direct $T_e$-based
abundances.  The empirical metallicity scale is likely the preferable
metallicity scale at present.

The construction of an empirical calibration encounters the following
difficulty. Not all direct abundances are of high precision since the
measurements of the weak auroral lines can involve considerable 
errors. Therefore the choice of a sample of H\,{\sc ii} regions
with reliable abundances is not a trivial task. We recently suggested
a new method (the ``C method'') for abundance determinations in
H\,{\sc ii} regions, which can be used over the whole range of
metallicities of H\,{\sc ii} regions and which provides oxygen and
nitrogen abundances on the same metallicity scale as the classical
$T_{e}$ method \citep{Pilyugin2012}.  It is important that the $C$
method allows one to choose a sample of H\,{\sc ii} regions with
reliable $T_{e}$-based abundances \citep{Pilyugin2012,Pilyugin2013,Zinchenko2016}. 

The goal of the present study is to establish simple calibration
relations that provide the oxygen and nitrogen abundance
determinations over the whole range of H\,{\sc ii} region
metallicities with relative errors less than 0.1 dex.  The 
paper is structured as follows.  The sample of calibrating data 
points is reported in Section 2.  The calibration relations are 
constructed in Section 3.  The discussion is given in Section 4, 
followed by a summary (Section 5).

Throughout the paper, we will use the following standard notations 
for the line intensities: \\ 
$R_2$  = $I_{\rm [O\,II] \lambda 3727+ \lambda 3729} /I_{{\rm H}\beta }$,  \\
$N_2$  = $I_{\rm [N\,II] \lambda 6548+ \lambda 6584} /I_{{\rm H}\beta }$,  \\
$S_2$  = $I_{\rm [S\,II] \lambda 6717+ \lambda 6731} /I_{{\rm H}\beta }$,  \\
$R_3$  = $I_{{\rm [O\,III]} \lambda 4959+ \lambda 5007} /I_{{\rm H}\beta }$.  \\

Based on these definitions, the excitation parameter $P$ is expressed
as $P$ = $R_{3}$/$R_{23}$ = $R_{3}$/($R_{2}$ + $R_{3}$).  The electron
temperatures will be given in units of 10$^4$K.  The notation
(O/H)$^{*}$ = 12 +log(O/H) will be used in order to permit us to write
our equations in a compact way.

\section{The sample of the calibrating data points}

\begin{figure}
\resizebox{1.00\hsize}{!}{\includegraphics[angle=000]{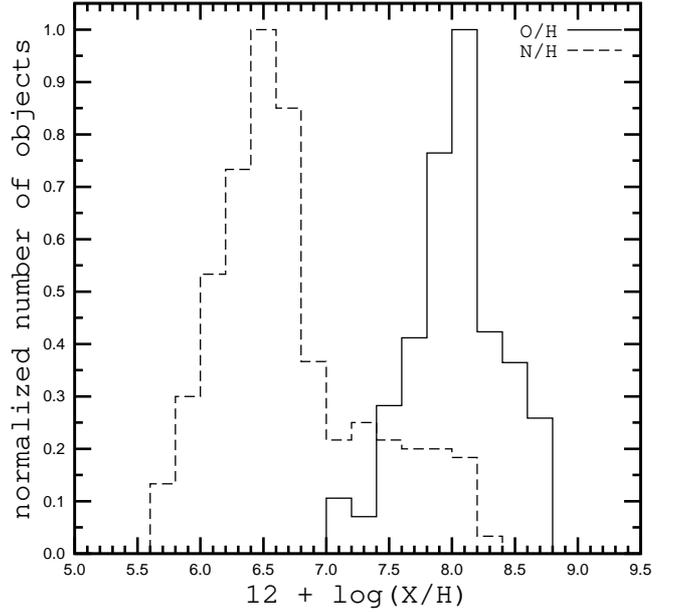}}
\caption{The normalized histograms of the oxygen and nitrogen abundances 
for our sample of calibrating H\,{\sc ii} regions. }
\label{figure:gistxhe}
\end{figure}

It was noted above that the choice of a sample of H\,{\sc ii} regions
with reliable abundance determinations is not a trivial task.  Such a
sample was compiled in the framework of the $C$ method for abundance
determinations in H\,{\sc ii} regions \citep{Pilyugin2012}. This
selection of reference H\,{\sc ii} regions is based on the idea that
if an H\,{\sc ii} region belongs to the sequence of photoionized
nebulae, and its line intensities are measured accurately, then
different methods based on different emission lines should yield
similar physical characteristics (such as electron temperatures and
abundances) for a given object \citep{Thuan2010,Pilyugin2012}. 

The original compilation of reference H\,{\sc ii} region spectra with
detected auroral lines (and, consequently, with $T_{e}$-based
abundances) that was created in \citet{Pilyugin2012} has been updated
by adding more recent measurements of H\,{\sc ii} regions.  The latest
version of the collection includes 965 $T_{e}$-based abundances.  Using
those data we select a sample of reference H\,{\sc ii} regions for
which all absolute differences in their oxygen abundances
(O/H)$_{C_{\rm ON}}$  -- (O/H)$_{T_{e}}$ and (O/H)$_{C_{\rm NS}}$  --
(O/H)$_{T_{e}}$ and nitrogen abundances (N/H)$_{C_{\rm ON}}$  --
(N/H)$_{T_{e}}$ and (N/H)$_{C_{\rm NS}}$  -- (N/H)$_{T_{e}}$ are less
than 0.1 dex. This sample of reference H\,{\sc ii} regions contains
313 objects \citep{Zinchenko2016} and will be used as calibrating data
set in the present study.  Hence, here we use the empirical
metallicity scale defined by H\,{\sc ii} regions with abundances
derived through the direct method ($T_e$ method).  

Fig.~\ref{figure:gistxhe} shows the normalized histograms of the
oxygen and nitrogen abundances for the H\,{\sc ii} regions of our
reference sample. 

\section{Calibration relations}

\subsection{Approach}

\begin{figure*}
\resizebox{0.980\hsize}{!}{\includegraphics[angle=000]{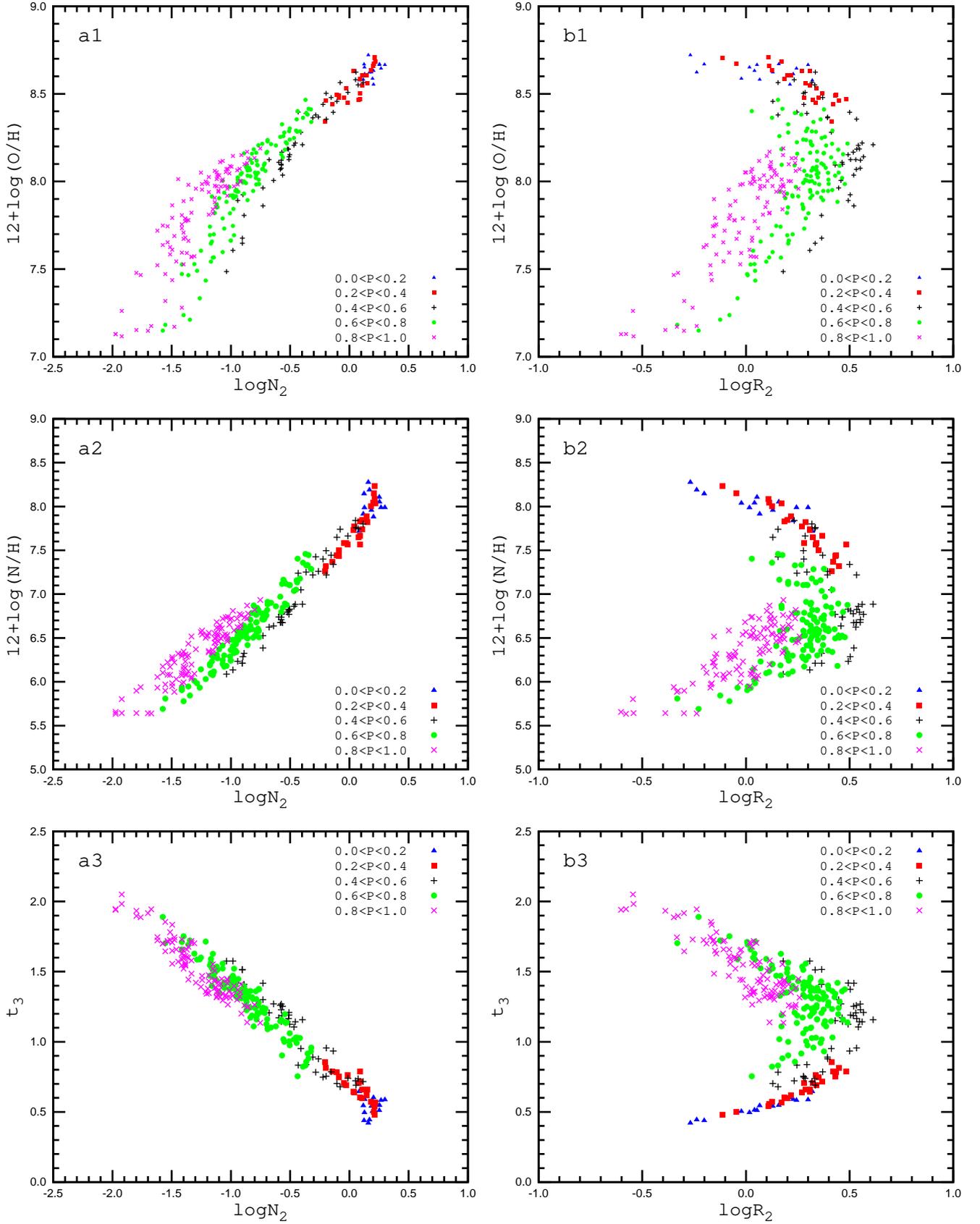}}
\caption{
Oxygen abundances (panel $a1$), nitrogen abundances (panel $a2$), 
and electron temperatures $t_{3}$ (panel $a3$) as a function
of the nitrogen line $N_{2}$ intensity for our sample of H\,{\sc ii}
regions used as calibrating data points.  H\,{\sc ii} regions with
different values of the excitation parameter $P$ are plotted with
different symbols as explained in the legend.  The panels $b1,
b2, b3$ show the same as the panels $a1, a2, a3$ but as a
function of the oxygen line $R_{2}$ intensity. 
}
\label{figure:ln2xhte}
\end{figure*}

The panels in the left column of Fig.~\ref{figure:ln2xhte} show the
oxygen abundances (panel $a1$), nitrogen abundances (panel $a2$), and
electron temperatures $t_{3}$ (panel $a3$) as a function of the
nitrogen line N$_{2}$ intensity for our sample of reference  H\,{\sc
ii} regions.  H\,{\sc ii} regions with different values of the
excitation parameter $P$ are plotted with different symbols.  It should be
noted that the electron temperature $t_{3}$ within the zone O$^{++}$
was not measured in some of the H\,{\sc ii} regions used as
calibrating data points.  Instead, the electron temperature $t_{2}$
within the zone N$^{+}$ or the electron temperature $t_{3,S}$ within
the zone S$^{++}$ was measured. In those cases, the electron
temperature $t_{3}$ was derived from the electron temperature $t_{2}$
adopting the commonly used relation between $t_{3}$ and $t_{2}$
\citep{Campbell1986,Garnett1992} or from the electron temperature
$t_{3,S}$ through the $t_{3}$ -- $t_{3,S}$ relation from
\citet{Garnett1992}.   

Panel $a3$ of Fig.~\ref{figure:ln2xhte} suggests that the nitrogen
line $N_{2}$ intensity can be used as an indicator of the electron
temperature in an H\,{\sc ii} region. The electron temperature $t_{3}$
is a monotonic function of logN${_2}$ across the whole range of
electron temperatures in  H\,{\sc ii} regions.  However, there is an
appreciable difference in this relation at high and low temperatures.
There is a distinct dependence of the $t$ -- $N_{2}$ relation on the
value of the excitation parameter $P$ at high electron temperatures
and this dependence disappears at low temperatures. The transition
between the two regimes happens at log$N_{2} \approx -0.6$.  The
nitrogen and oxygen abundances are also a monotonic functions of the
nitrogen line N$_{2}$ intensity as can be seen in panels $a2$ and $a1$
of Fig.~\ref{figure:ln2xhte}. The  N/H -- $N_{2}$ (or O/H -- $N_{2}$)
relation depends also on the value of the excitation parameter $P$ at
low metallicities (high electron temperatures) and this dependence
disappears at high metallicities (low electron temperatures). 

The variation of the $t$ -- $N_{2}$ and the X/H -- $N_{2}$ relations
with the value of the excitation parameter $P$ seems to reflect the
change of the contribution of the emission of the low ionization zone
to the total emission of an H\,{\sc ii} region.  If this is the case
then the difference in the dependence of the $t$ -- $N_{2}$ and the
X/H -- $N_{2}$ relations on the value of the excitation parameter $P$
at high and low electron temperatures can be interpreted in the
following way.  Hot H\,{\sc ii} regions have usually a high excitation
level while cool H\,{\sc ii} regions are areas of low excitation. A
similar change of the value of the excitation parameter $P$ (say, by
0.1) corresponds to different changes of the low ionization zone
contribution to the total emission in hot and cool H\,{\sc ii}
regions.  Indeed, the variation of $P$ from 0.8 to 0.9 in hot H\,{\sc
ii} regions corresponds to the change of the low ionization zone
contribution by a factor of 2 (from 20 to 10\%) while the similar
variation of $P$ from 0.1 to 0.2 in cool H\,{\sc ii} regions
corresponds to the change of the low ionization zone contribution by
a factor of only $\approx 1.12$ (from 90 to 80\%). 

The panels in the right column of Fig.~\ref{figure:ln2xhte} show the
oxygen abundance (panel $b1$), the nitrogen abundance (panel $b2$),
and the electron temperature $t_{3}$ (panel $b3$) as a function of
oxygen line R$_{2}$ intensity for our calibration sample.  H\,{\sc
ii} regions with different values of the excitation parameter $P$ are
shown by different symbols.  It is well known that the relation
between the oxygen abundance and the strong oxygen line intensities is
double-valued with two distinct parts traditionally labeled as the
upper (high-metallicity) and lower (low-metallicity) branch.  The
panels $b2$ and $b3$ of Fig.~\ref{figure:ln2xhte} show that the
general behaviour of the N/H -- $R_{2}$ and the $t$ -- $R_{2}$
diagrams are similar to that of the  O/H -- $R_{2}$ diagram, i.e.,
they are also double-valued.  Again the $t$ -- $R_{2}$ (and X/H --
$R_{2}$) relation depends on the value of the excitation parameter
$P$ at low metallicities (high electron temperatures). This dependence
disappears at high metallicities (low electron temperatures). 

Thus, the examination of  Fig.~\ref{figure:ln2xhte} suggests that
different O/H --  $R_{2}$ (and N/H --  $R_{2}$) relations should be
constructed for the upper and lower branches.  The transition from the
upper to lower branch occurs at log$N_{2} \sim-0.6$.  It should be
noted that the use of the fixed value of $N_{2}$ as a dividing
criterion between the H\,{\sc ii} regions of the upper and lower
branches may be an oversimplification.  A more sophisticated condition
such as a fixed electron temperature or/and a fixed oxygen abundance
may provide a more reliable boundary criterion between objects on the
lower and upper branches. 

In general, the relation between the oxygen abundance in an H\,{\sc
ii} region on the upper (lower) branch and the oxygen $R_{2}$ line
intensity in its spectrum depends on two parameters: the electron
temperature and the level of excitation.  We consider the simplest
linear expression   
\begin{equation}
12 + \log ({\rm O/H})  = a(t,P) + b(t,P) \, \log R_{2} 
\label{equation:ohfunc}
\end{equation}
where the coefficients $a$ and $b$ depend on the electron temperature
and the excitation parameter.  In this study we will use the value of
log($R_{3}/R_{2}$) as an indicator of the excitation level of an
H\,{\sc ii} region and the value of the log$N_{2}$ as an index of its
electron temperature.  Under the assumption that the coefficients $a$
and $b$ depend linearly on the electron temperature $t$ and the
excitation parameter $P$, Eq.~\ref{equation:ohfunc} can be
rewritten as    
\begin{eqnarray}
       \begin{array}{lll}
{\rm (O/H)}^{*}       & = &   a_{1} + a_{2} \, \log (R_{3}/R_{2}) + a_{3} \, \log N_{2}  \\  
                     & + &  (a_{4} + a_{5} \, \log (R_{3}/R_{2}) + a_{6} \, \log N_{2}) \times \log R_{2}   \\ 
     \end{array}
\label{equation:ohgeneral}
\end{eqnarray}
where the notation (O/H)$^{*} \equiv$ 12 +log(O/H) is used for the
sake of the compact writing of the equation. We adopt an expression of
this form as the calibration relation for the oxygen abundance
determinations.  Thus, we consider the three-dimensional relation
O/H=$f$($R_{2}$,$R_{3}$,$N_{2}$).  A comparison between the panels $b1$
and $b2$ of Fig.~\ref{figure:ln2xhte} suggests that an expression of
similar form can also be used as a calibration relation for the
nitrogen abundance determinations.

\subsection{Calibration relations for determinations of the oxygen abundance} 

\subsubsection{R calibrations} 

\begin{figure*}
\resizebox{1.00\hsize}{!}{\includegraphics[angle=000]{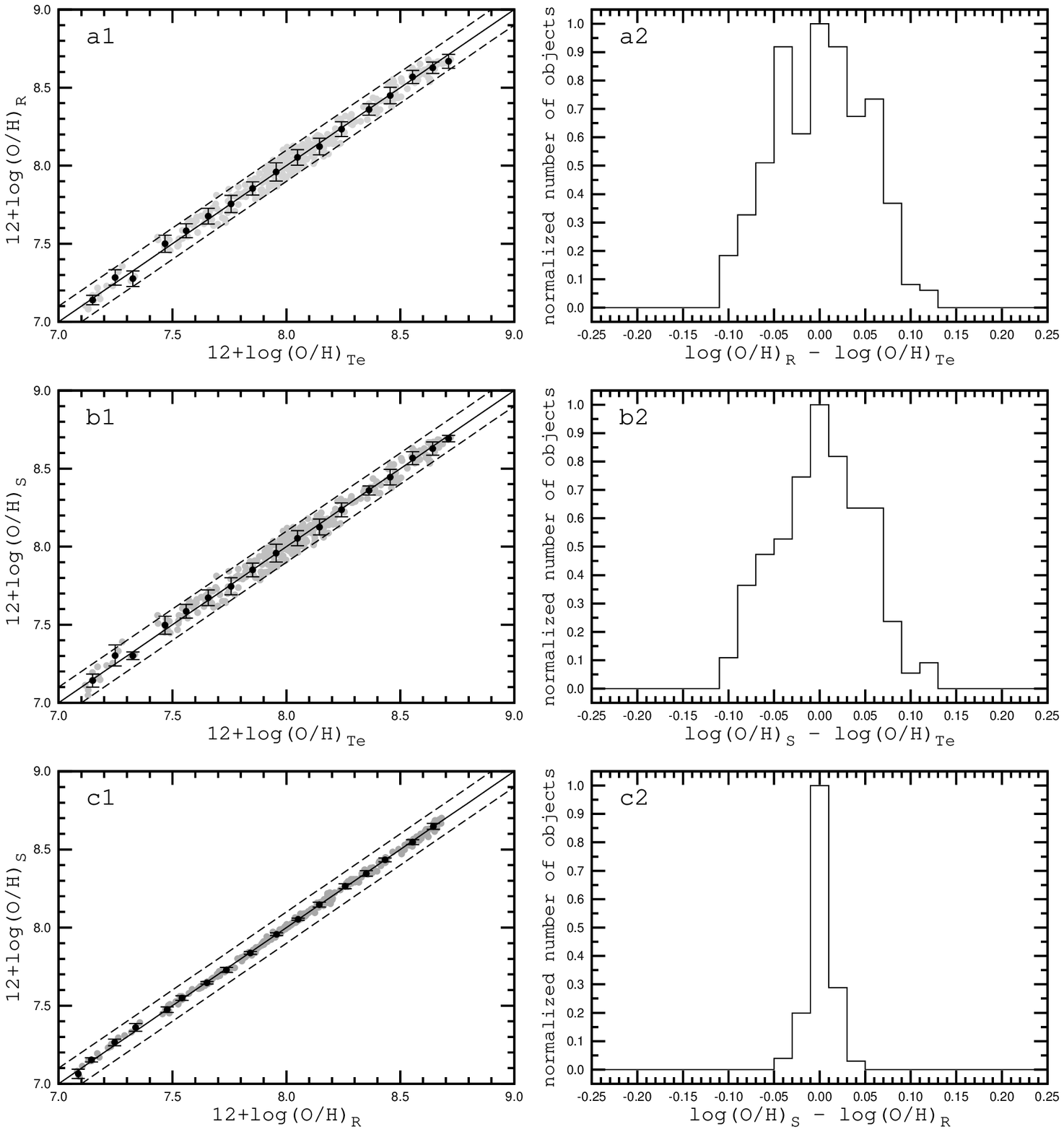}}
\caption{
Panel $a1$ shows the oxygen abundance (O/H)$_{R}$ as a function of
oxygen abundance (O/H)$_{T_{e}}$ for the calibrating H\,{\sc ii}
regions (313 data points).  The grey points stand for individual
H\,{\sc ii} regions.  The dark points represent the mean values of the
abundances in bins of 0.1 dex in (O/H)$_{T_{e}}$.  The bars are the 
mean values of the differences of the oxygen abundances (O/H)$_{R}$ and
(O/H)$_{T_{e}}$ of our H\,{\sc ii} regions in bins.  The solid line is
that of equal values; the dashed lines show the $\pm 0.1$ dex
deviations from unity.  Panel $a2$ shows the normalized histogram of
the differences between the (O/H)$_{R}$ and (O/H)$_{T_{e}}$
abundances.  The panels $b1$ and $b2$ are the same as the panels
$a1$ and $a2$ but for the abundances (O/H)$_{S}$ and (O/H)$_{T_{e}}$. 
The panels $c1$ and $c2$ are the same as the panels $a$ and $a2$ but 
for the abundances (O/H)$_{S}$ and (O/H)$_{R}$.}
\label{figure:ohteoh}
\end{figure*}

\begin{figure*}
\resizebox{1.00\hsize}{!}{\includegraphics[angle=000]{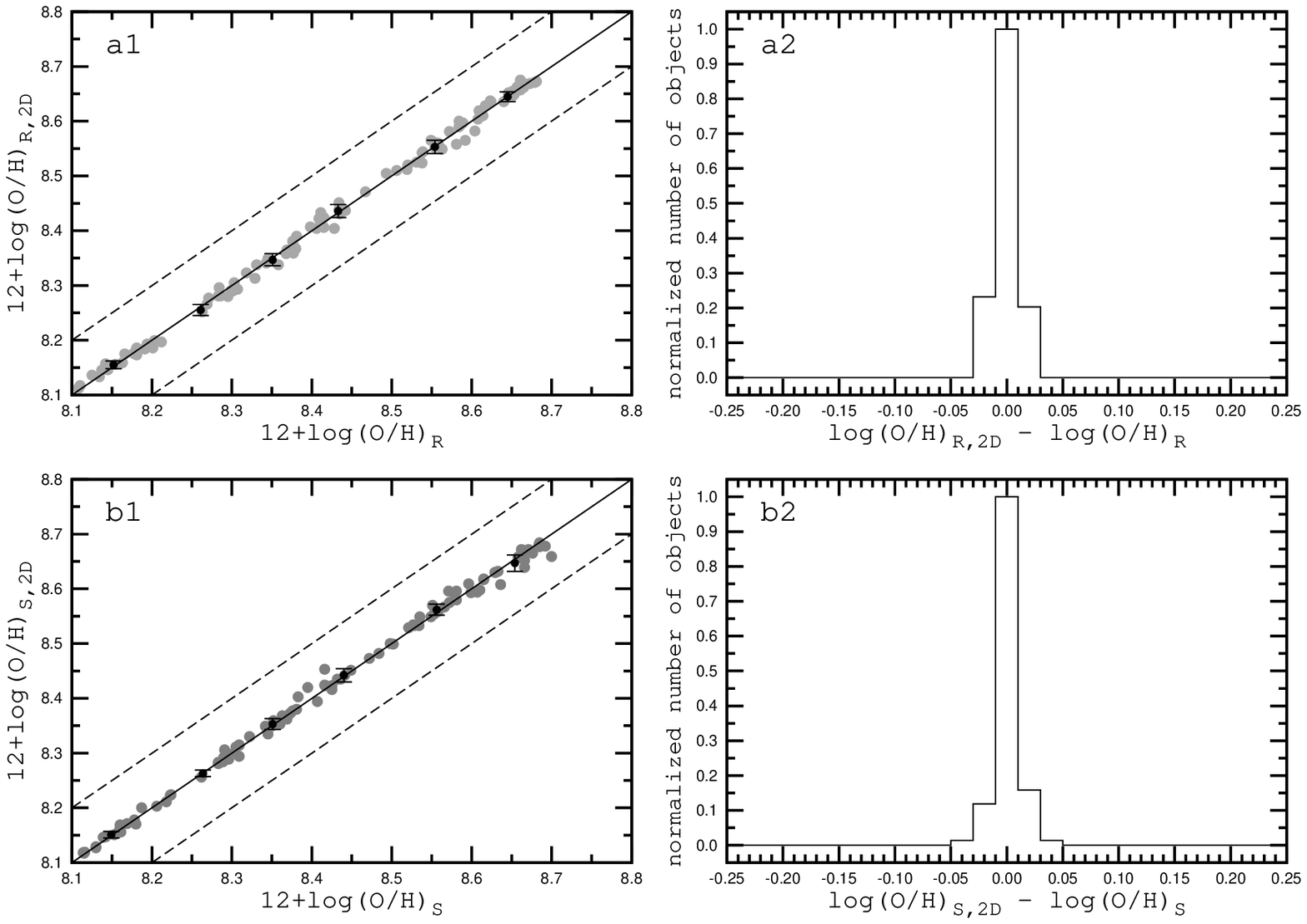}}
\caption{Panel $a1$ shows the oxygen abundance (O/H)$_{R,2D}$ as a
function of the oxygen abundance (O/H)$_{R}$ for the calibrating
H\,{\sc ii} regions.  The grey points depict the individual H\,{\sc
ii} regions.  The dark points represent the mean values of the oxygen
abundances in bins of 0.1 dex in (O/H)$_{R}$.  The bars are the mean
values of the differences between the oxygen abundances (O/H)$_{R,2D}$
and (O/H)$_{R}$ of our H\,{\sc ii} regions in bins.  The solid line 
that of equal values.  The dashed lines show the $\pm$0.1 dex deviation
represents equality.  Panel $a2$ shows the normalized histogram of
the differences between the (O/H)$_{R,2D}$ and (O/H)$_{R}$ abundances.
The panels $b1$ and $b2$ show the same as the panels $a1$ and $a2$ but
for the abundances (O/H)$_{S,2D}$ and (O/H)$_{S}$. }
\label{figure:ohteoh2d}
\end{figure*}

First we consider the calibration for abundance determinations in
H\,{\sc ii} regions for the case when the $R_{2}$ line is available.  A
calibration of this type will be referred to as $R$ calibration.  The
oxygen abundance determined using such a calibration will be labeled as
(O/H)$_{R}$.

The set of coefficients $a_{j}$ in Eq.~\ref{equation:ohgeneral} can be
derived by imposing the condition that the average value of the
differences between the oxygen abundances determined through the
calibration and the original ones, 
\begin{equation}
[(O/H)^{*}_{R} - (O/H)_{T_{e}}^{*}]^{mean} = \left(\frac{1}{n}\sum_{j=1}^{n} [(O/H)^{*}_{R,j} - (O/H)^{*}_{T_{e},j}]^{2}\right)^{1/2}  
\label{equation:meandif}
\end{equation}
is minimzed for our sample of calibrating data points. The notation
(O/H)$^{*}$ $\equiv$ 12 +log(O/H) is used to allow us to write the
equation in a more compact manner. 

For H\,{\sc ii} regions with log$N_{2} \ge -0.6$  (the upper
branch), the following values of the coefficients were obtained: 
      $a_{1} =  8.589$,  
      $a_{2} =  0.022$,
      $a_{3} =  0.399$, 
      $a_{4} = -0.137$, 
      $a_{5} =  0.164$,
      $a_{6} =  0.589$. 
This then results in
\begin{eqnarray}
       \begin{array}{lll}
     {\rm (O/H)}^{*}_{R,U}  & = &   8.589 + 0.022 \, \log (R_{3}/R_{2}) + 0.399 \, \log N_{2}   \\  
                          & + &  (-0.137 + 0.164 \, \log (R_{3}/R_{2}) + 0.589 \log N_{2})   \\ 
                          & \times &  \log R_{2}   \\ 
     \end{array}
\label{equation:ohru}
\end{eqnarray}
where (O/H)$^{*}_{R,U}$ = 12 +log(O/H)$_{R,U}$. 

For H\,{\sc ii} regions with log$N_{2} < -0.6$ (the lower branch), 
the obtained values of the coefficients are: 
      $a_{1} =  7.932$,  
      $a_{2} =  0.944$, 
      $a_{3} =  0.695$, 
      $a_{4} =  0.970$, 
      $a_{5} = -0.291$, 
      $a_{6} = -0.019$, 
and 
\begin{eqnarray}
       \begin{array}{lll}
     {\rm (O/H)}^{*}_{R,L}  & = &   7.932 + 0.944 \, \log (R_{3}/R_{2}) + 0.695 \, \log N_{2}   \\  
                          & + &  (0.970 - 0.291 \, \log (R_{3}/R_{2}) - 0.019 \log N_{2})   \\ 
                          & \times & \log R_{2}   \\ 
     \end{array}
\label{equation:ohrl}
\end{eqnarray}
where (O/H)$^{*}_{R,L}$ = 12 +log(O/H)$_{R,L}$. 

Panel $a1$ of Fig.~\ref{figure:ohteoh} shows the oxygen abundance
(O/H)$_{R}$ as a function of the oxygen abundance (O/H)$_{T_{e}}$  for
the calibrating data points. The grey points mark individual H\,{\sc
ii} regions. We split our sample of H\,{\sc ii} regions into bins of
0.1 dex in the (O/H)$_{T_{e}}$ abundance. The mean oxygen abundances
(O/H)$_{T_{e}}^{mean}$ and (O/H)$_{R}^{mean}$  for the H\,{\sc ii}
regions in each bin were determined. The absolute values of the mean
difference between (O/H)$_{R}$ and (O/H)$_{T_{e}}$ were estimated for
each bin using Eq.~\ref{equation:meandif}.  The mean values of
(O/H)$_{T_{e}}^{mean}$ and (O/H)$_{R}^{mean}$ are shown in panel
$a1$ of Fig.~\ref{figure:ohteoh} as dark points while the bars
show the absolute values of the mean difference between the oxygen
abundances (O/H)$_{R}$ and (O/H)$_{T_{e}}$ for each bin.  The solid
line is that of equal values; the dashed lines show the $\pm 0.1$ dex
deviations from the unity line.   

Panel $a2$ of Fig.~\ref{figure:ohteoh} shows the normalized histogram
of the differences between the calibration-based oxygen abundances
(O/H)$_{R}$ inferred here and the directly measured 
(O/H)$_{T_{e}}$ values. 

The examination of panels $a1$ and $a2$ of Fig.~\ref{figure:ohteoh}
shows that the oxygen abundances (O/H)$_{R}$ and (O/H)$_{T_{e}}$ agree
usually within 0.1 dex.  The mean difference for the 313 calibrating
H\,{\sc ii} regions is 0.049 dex.  On the one hand, this suggests that
our reference sample does indeed contain H\,{\sc ii} regions with
reliable oxygen abundances (O/H)$_{T_{e}}$.  On the other hand, this
indicates that the chosen form of the calibration relations,
Eq.~\ref{equation:ohgeneral}, allows us to reproduce the relation
between the oxygen abundances in an H\,{\sc ii} region and the
intensities of the strong emission lines in its spectrum rather well.

\subsubsection{S calibrations} 

There are many measurements of H\,{\sc ii} regions where the line
$R_{2}$ is not available, for instance in the spectra of nearby
galaxies in the Sloan Digital Sky Survey (SDSS). It has been argued
that the oxygen abundance in an H\,{\sc ii} region can be estimated
even if the line $R_{2}$ is not available \citep{Pilyugin2011}. Now we
consider the calibration for such a case.  We will use the sulphur line
$S_{2}$ intensity instead of the oxygen $R_{2}$ line intensity.  A
calibration of this type will be referred to as $S$ calibration, and
the oxygen abundance determined using this kind of calibration will be
labeled (O/H)$_{S}$.  The same form of the calibration relation as
in Eq.~\ref{equation:ohgeneral} is adopted. 

Using the calibrating H\,{\sc ii} regions of the upper branch
(log$N_{2} \ge -0.6$), the following values of the coefficients
$a_{j}$ were obtained: 
      $a_{1} =  8.424$,  
      $a_{2} =  0.030$,
      $a_{3} =  0.751$, 
      $a_{4} = -0.349$, 
      $a_{5} =  0.182$,
      $a_{6} =  0.508$. 
The corresponding calibration relation is  
\begin{eqnarray}
       \begin{array}{lll}
     {\rm (O/H)}^{*}_{S,U}  & = &   8.424 + 0.030 \, \log (R_{3}/S_{2}) + 0.751 \, \log N_{2}   \\  
                          & + &  (-0.349 + 0.182 \, \log (R_{3}/S_{2}) + 0.508 \log N_{2})   \\ 
                          & \times & \log S_{2}   \\ 
     \end{array}
\label{equation:ohsu}
\end{eqnarray}
where (O/H)$^{*}_{S,U}$ = 12 +log(O/H)$_{S,U}$. 
Using the calibrating H\,{\sc ii} regions of the lower branch 
(log$N_{2} < -0.6$), we obtained the following values for the 
coefficients $a_{j}$: 
      $a_{1} =  8.072$,  
      $a_{2} =  0.789$, 
      $a_{3} =  0.726$, 
      $a_{4} =  1.069$, 
      $a_{5} = -0.170$, 
      $a_{6} =  0.022$. 
Then
\begin{eqnarray}
       \begin{array}{lll}
     {\rm (O/H)}^{*}_{S,L}  & = &   8.072 + 0.789 \, \log (R_{3}/S_{2}) + 0.726 \, \log N_{2}   \\  
                          & + &  (1.069 - 0.170 \, \log (R_{3}/S_{2}) + 0.022 \log N_{2})    \\ 
                          & \times & \log S_{2}   \\ 
     \end{array}
\label{equation:ohsl}
\end{eqnarray}
where (O/H)$^{*}_{S,L}$ = 12 +log(O/H)$_{S,L}$. 

Panel $b1$ of Fig.~\ref{figure:ohteoh} shows the comparison between
the oxygen abundances (O/H)$_{S}$ and (O/H)$_{T_{e}}$ for our sample
of reference  H\,{\sc ii} regions.  The grey  points denote the values
for the individual H\,{\sc ii} regions.  The dark points represent the
mean oxygen abundances (O/H)$_{T_{e}}^{mean}$ and (O/H)$_{S}^{mean}$
for the H\,{\sc ii} regions in bins of 0.1 dex in (O/H)$_{T_{e}}$.  The
bars show the absolute values of the mean difference between the
oxygen abundances (O/H)$_{S}$ and (O/H)$_{T_{e}}$ for each bin.  The
solid line indicates equality.  The dashed lines show the $\pm$0.1 dex
deviation from these equal values.  In panel $b2$ of
Fig.~\ref{figure:ohteoh} we plot the normalized histogram of the
differences between the calibration-based oxygen abundances
(O/H)$_{S}$ and the directly measued (O/H)$_{T_{e}}$ values.
Inspection of panels $b1$ and $b2$ of Fig.~\ref{figure:ohteoh} shows
that the difference between the oxygen abundances (O/H)$_{S}$ and
(O/H)$_{T_{e}}$ is usually less than 0.1 dex.  The mean difference for
our 313 calibrating H\,{\sc ii} regions is 0.048 dex. 

Panel $c1$ of Fig.~\ref{figure:ohteoh} shows the comparison between
the oxygen abundances (O/H)$_{S}$ and (O/H)$_{R}$ for our calibrating
data points.  Panel $c2$ of Fig.~\ref{figure:ohteoh} shows the
normalized histogram of the differences between the oxygen abundances
(O/H)$_{S}$ and (O/H)$_{R}$.  The panels $c1$ and $c2$ of
Fig.~\ref{figure:ohteoh} demonstrate that the $R$ and $S$ calibrations
produce oxygen abundance values that are very close to each other.
The mean difference for the 313 calibrating H\,{\sc ii} regions
amounts to only 0.013 dex.

\subsubsection{Two-dimensional $R$ and $S$ calibrations for the upper branch} 

The calibration relations suggested above are three-dimensional, i.e.,
the calibration for the oxygen abundance determination involves three
parameters: log($R_{3}/R_{2}$), log$N_{2}$, and log$R_{2}$, or
log($R_{3}/S_{2}$), log$N_{2}$, and log$S_{2}$. It was noted above
(see Fig.~\ref{figure:ln2xhte}) that there is a distinct dependence of
the oxygen abundance on the value of the excitation parameter $P$ at
high electron temperatures (lower branch) and this dependence
disappears at low temperatures (upper branch).  Hence one may expect
that the dependence of the oxygen abundance on the value of the
excitation parameter $P$ can be neglected and two-dimensional
calibrations can be determined for the upper branch.  

The obtained two-dimensional $R$ relation for the upper branch is 
\begin{eqnarray}
       \begin{array}{lll}
12+{\rm (O/H)}_{R,2D}  & = &  8.589 + 0.329  \, \log N_{2}  \\  
                      & + & (-0.205 + 0.549 \, \log N_{2}) \times \log R_{2}  \\ 
     \end{array}
\label{equation:ohr1p}
\end{eqnarray}
The inferred two-dimensional $S$ relation for the upper branch is 
\begin{eqnarray}
       \begin{array}{lll}
12+{\rm (O/H)}_{S,2D}  & = &  8.445 + 0.699  \, \log N_{2}  \\  
                      & + & (-0.253 + 0.217 \, \log N_{2}) \times \log S_{2}  \\ 
     \end{array}
\label{equation:ohs1p}
\end{eqnarray}

Panel $a1$ of Fig.~\ref{figure:ohteoh2d} shows the comparison between
oxygen abundances for the calibrating H\,{\sc ii} regions obtained
through the three-dimensional calibration (O/H)$_{R}$ and through the
two-dimensional calibration (O/H)$_{R,2D}$.  Panel $a2$ of
Fig.~\ref{figure:ohteoh2d} displays the normalized histogram of the
differences between the oxygen abundances (O/H)$_{R,2D}$ and
(O/H)$_{R}$.  The panels $b1$ and $b2$ of Fig.~\ref{figure:ohteoh2d} show
a similar comparison between the oxygen abundances (O/H)$_{S}$ and
(O/H)$_{S,2D}$.  Fig.~\ref{figure:ohteoh2d} demonstrates that
the two-dimensional and three-dimensional $R$ and $S$ calibrations
produce oxygen abundances for H\,{\sc ii} regions of the upper branch
that are in very good agreement with one another.

Thus, the oxygen abundances in high-metallicity H\,{\sc ii} regions
(the upper branch) can be estimated using the intensities of only two
lines, namely $R_{2}$ and $N_{2}$ (or $S_{2}$ and $N_{2}$).

\begin{figure*}
\resizebox{1.00\hsize}{!}{\includegraphics[angle=000]{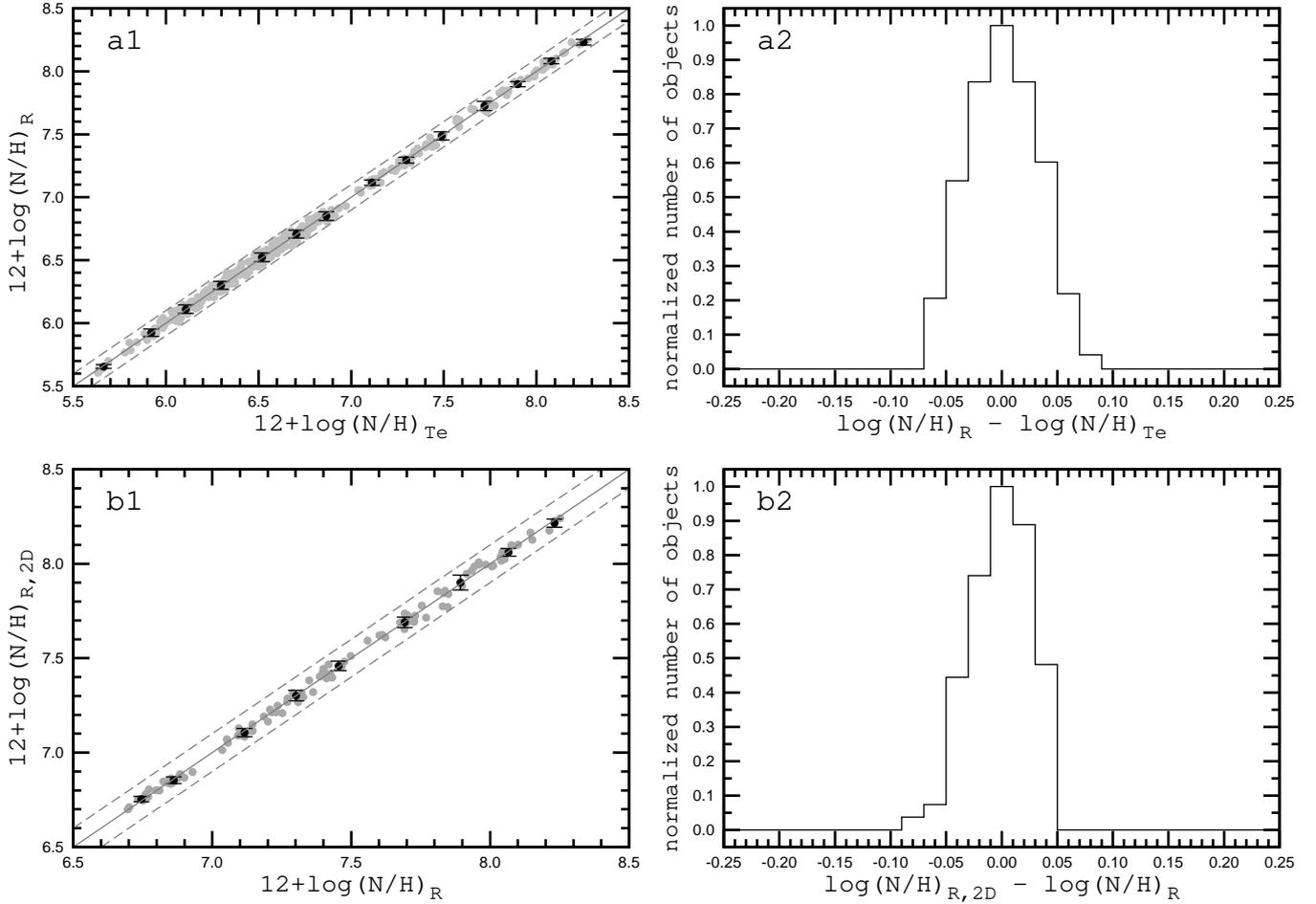}}
\caption{Panel $a1$ shows the nitrogen abundance (N/H)$_{R}$ as a
function of the nitrogen abundance (N/H)$_{T_{e}}$ for the calibrating
H\,{\sc ii} regions.  The grey points mark individual H\,{\sc ii}
regions.  The dark points represent the mean values of the abundances
in bins of 0.1 dex in (N/H)$_{T_{e}}$.  The bars denote the mean
values of the differences of the oxygen abundances (N/H)$_{R}$ and
(N/H)$_{T_{e}}$ of the H\,{\sc ii} regions in bins.  The solid line
indicates equal values; the dashed lines show the $\pm$0.1 dex
deviation from unity.  Panel $a2$ contains the normalized histogram of
the differences between the (N/H)$_{R}$ and (N/H)$_{T_{e}}$
abundances.  The panels $b1$ and $b2$ are the same as the panels $a1$
and $a2$ but for the abundances (N/H)$_{R,2D}$ and (N/H)$_{R}$. }
\label{figure:nhtenh}
\end{figure*}

\subsection{Calibration relations for determinations of the nitrogen abundance} 

Here we establish calibration relations for nitrogen abundance
determinations in H\,{\sc ii} regions.  Fig.~\ref{figure:ln2xhte}
shows that the general behaviour of the nitrogen abundances in the
considered diagrams is similar to that of the oxygen abundances. This
suggests that an expression of the same form as
Eq.~\ref{equation:ohgeneral} can be adopted for the calibration
relation for the nitrogen abundance determinations. 

Based on the H\,{\sc ii} regions of the upper branch 
(log$N_{2} \ge -0.6$), the following values of the coefficients 
$a_j$ were obtained: 
      $a_{1} =  7.939$,  
      $a_{2} =  0.135$,
      $a_{3} =  1.217$, 
      $a_{4} = -0.765$, 
      $a_{5} =  0.166$,
      $a_{6} =  0.449$. 
The resulting calibration relation for the nitrogen abundance 
determination is
\begin{eqnarray}
       \begin{array}{lll}
     {\rm (N/H)}^{*}_{R,U}  & = &   7.939 + 0.135 \, \log (R_{3}/R_{2}) + 1.217 \, \log N_{2}   \\  
                          & + &  (-0.765 + 0.166 \, \log (R_{3}/R_{2}) + 0.449 \log N_{2})   \\ 
                          & \times & \log R_{2}   \\ 
     \end{array}
\label{equation:nhru}
\end{eqnarray}
where (N/H)$^{*}_{R,U}$ = 12 +log(N/H)$_{R,U}$. 

For the H\,{\sc ii} regions on the lower branch (log$N_{2} < -0.6$), 
the values of the coefficients are: 
      $a_{1} =  7.476$,  
      $a_{2} =  0.879$,
      $a_{3} =  1.451$, 
      $a_{4} = -0.011$, 
      $a_{5} = -0.327$,
      $a_{6} = -0.064$,
and the calibration relation is 
\begin{eqnarray}
       \begin{array}{lll}
     {\rm (N/H)}^{*}_{R,L}  & = &   7.476 + 0.879 \, \log (R_{3}/R_{2}) + 1.451 \, \log N_{2}   \\  
                          & + &  (-0.011 - 0.327 \, \log (R_{3}/R_{2}) - 0.064 \log N_{2})   \\ 
                          & \times & \log R_{2}   \\ 
     \end{array}
\label{equation:nhrl}
\end{eqnarray}
where (N/H)$^{*}_{R,L}$ = 12 +log(N/H)$_{R,L}$.

Panel $a1$ of Fig.~\ref{figure:nhtenh} shows the comparison between
the nitrogen abundances (N/H)$_{R}$ and (N/H)$_{T_{e}}$ for our sample
of calibrating H\,{\sc ii} regions.  The grey points denote the values
for the individual H\,{\sc ii} regions.  The dark points represent the
mean nitrogen abundances (N/H)$_{T_{e}}^{mean}$ and (N/H)$_{R}^{mean}$
for the H\,{\sc ii} regions in bins of 0.2 dex in (N/H)$_{T_{e}}$.  The
bars show the absolute values of the mean difference between the
nitrogen abundances (N/H)$_{R}$ and (N/H)$_{T_{e}}$ for each bin.  The
solid line is that of equal values; the dashed lines delineate the
$\pm$0.1 dex deviation from equality.  Panel $a2$ of
Fig.~\ref{figure:nhtenh} shows the normalized histogram of the
differences between the calibration-based nitrogen abundances
(N/H)$_{R}$ and the directly measured (N/H)$_{T_{e}}$ values.  The
panels $a1$ and $a2$ of Fig.~\ref{figure:nhtenh} demonstrate that the
difference between the nitrogen abundances (N/H)$_{R}$ and
(N/H)$_{T_{e}}$ is usually less than 0.1 dex.  The mean difference for
the 313 calibrating H\,{\sc ii} regions is 0.031 dex. 

A comparison of the panels $a1$ and $a2$ of Fig.~\ref{figure:ohteoh}
with the panels $a1$ and $a2$ of Fig.~\ref{figure:nhtenh} shows that
the agreement between the $R$-calibration-based and the $T_{e}$-based
nitrogen abundances is better than that for the oxygen abundances.
Indeed, the mean difference (N/H)$_{R}$ -- (N/H)$_{T_{e}}$ is 0.031
dex while the mean difference (O/H)$_{R}$ -- (O/H)$_{T_{e}}$ is 0.049
dex.  

We also constructed the $S$ calibration for the determination of the
nitrogen abundances. However, the nitrogen abundances estimated with
the $S$ calibration are much more uncertain those produced by the $R$
calibration. Therefore we do not discuss the $S$ calibration for
nitrogen abundance determinations here. 

As in the case of oxygen, the two-dimensional $R$ calibration was
constructed for the high-metallicity  H\,{\sc ii} regions (upper
branch).  We obtained the relation  
\begin{eqnarray}
       \begin{array}{lll}
12+{\rm (N/H)}_{R,2D}  & = &  7.903 + 0.927  \, \log N_{2}  \\  
                      & + & (-0.819 + 0.686 \, \log N_{2}) \times \log(R_{2})  \\ 
     \end{array}
\label{equation:nh1p}
\end{eqnarray}
Panel $b1$ of Fig.~\ref{figure:nhtenh} shows the comparison between
the nitrogen abundances of the calibrating H\,{\sc ii} regions
obtained through the three-dimensional calibration (N/H)$_{R}$ and
through the two-dimensional calibration (N/H)$_{R,2D}$.  Panel $b2$ of
Fig.~\ref{figure:nhtenh} displays the normalized histogram of the
differences between the nitrogen abundances (N/H)$_{R,2D}$ and
(N/H)$_{R}$.  The comparison of the panels $a1$ and $a2$ of
Fig.~\ref{figure:ohteoh2d} with the panels $b1$ and $b2$ of
Fig.~\ref{figure:nhtenh} shows that the abundances produced by the
three- and two-dimensional $R$ calibrations agree better for oxygen
than for nitrogen.

Thus, the nitrogen abundance in an H\,{\sc ii} region can be obtained
from the intensities of the strong emission lines $R_{2}$, $R_{3}$,
and $N_{2}$ in its spectrum. Those abundances agree with the directly
measured nitrogen abundances within $\sim 0.1$ dex. The nitrogen
abundances in high-metallicity H\,{\sc ii} regions (the upper branch)
can also be estimated using the intensities of the two strong lines
$R_{2}$ and $N_{2}$ only.

\subsection{Calibration relation for determinations of the N/O ratio}

\begin{figure*}
\resizebox{1.00\hsize}{!}{\includegraphics[angle=000]{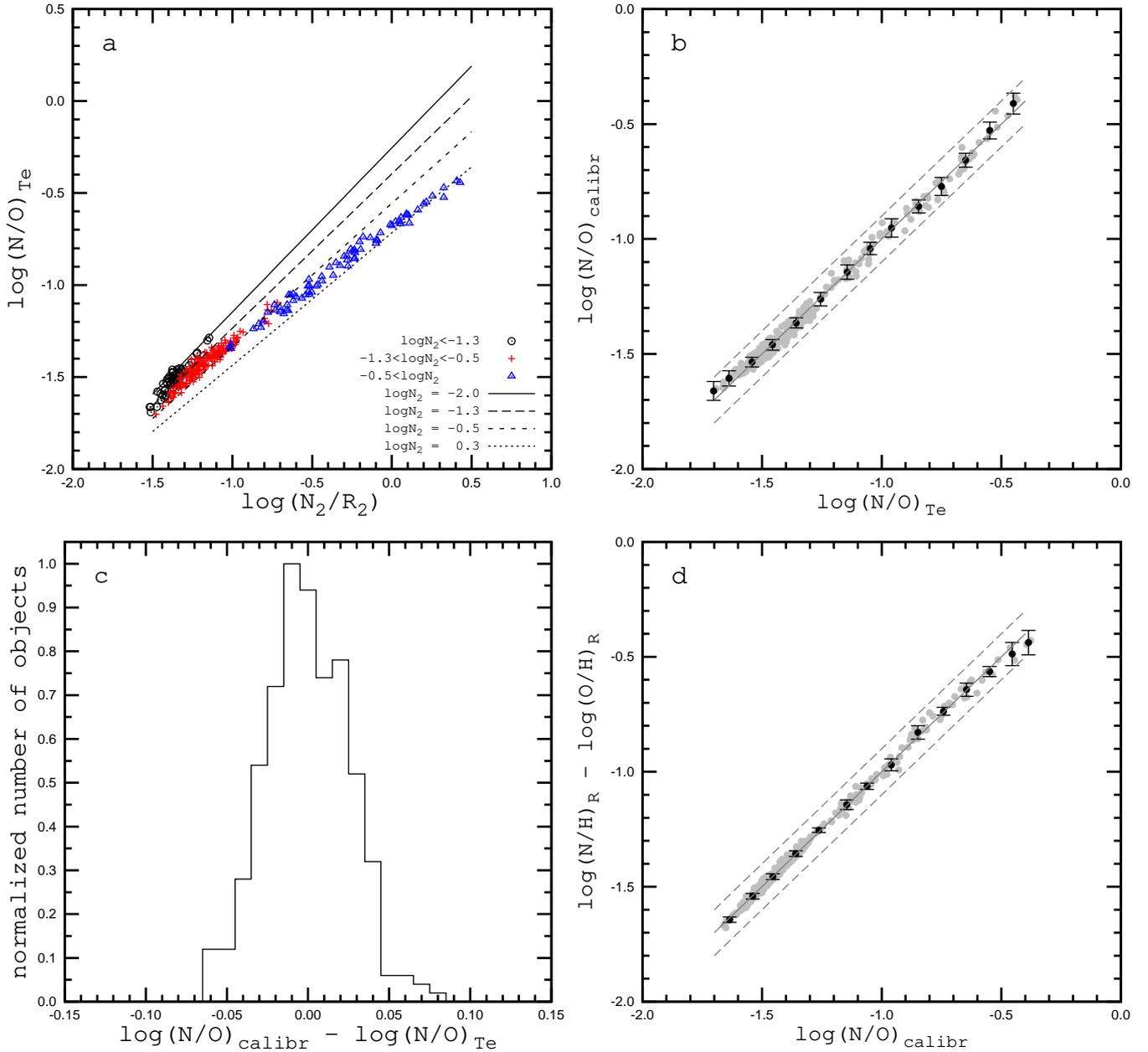}}
\caption{
Panel $a$.  The nitrogen-to-oxygen abundance ratio N/O as a
function of the line intensity ratio N$_{2}$/R$_{2}$ for the
calibrating H\,{\sc ii} regions. H\,{\sc ii} regions with $N_{2}$
intensities in three different intervals are shown by different
symbols.  The solid, dashed, and dotted lines show the calibration
relation for different values of the intensity of the line $N_{2}$.
Panel $b$.  The nitrogen-to-oxygen abundance ratio (N/O)$_{calibr}$ 
derived from the calibration relation (Eq.~\ref{equation:nolin})
as a function of nitrogen-to-oxygen abundance ratio (N/O)$_{T_{e}}$ 
for the calibrating H\,{\sc ii} regions (313 data points).  
The grey points stand for individual H\,{\sc ii} regions. 
The dark points represent the mean values of the 
nitrogen-to-oxygen abundance ratios in bins of 0.1 dex in (N/O)$_{T_{e}}$. 
The bars are mean values of the differences of abundance ratios (N/O)$_{calibr}$ 
and (N/O)$_{T_{e}}$ of our H\,{\sc ii} regions in bins.  The solid line is
that of equal values; the dashed lines show the $\pm$0.1 dex
deviations from unity.
Panel $c$.  Normalized histogram of the differences between N/O
ratios derived from the calibration relation
(Eq.~\ref{equation:nolin}) and through the $T_{e}$ method.  
Panel $d$.  The same as the panel $b$ but for nitrogen-to-oxygen 
abundance ratios (N/O)$_{calibr}$ derived from the NO calibration relation 
(Eq.~\ref{equation:nolin}) and nitrogen-to-oxygen abundance ratios 
determined as log(N/O)$_{R}$ = log(N/H)$_{R}$ --  log(O/H)$_{R}$ where nitrogen   
(N/H)$_{R}$ and oxygen (O/H)$_{R}$ abundances are determined separately from 
the corresponding calibration relations 
(Eqs.~\ref{equation:ohru}, \ref{equation:ohrl}, \ref{equation:nhru}, \ref{equation:nhrl}). 
}
\label{figure:nolin}
\end{figure*}

It is interesting to also establish the expression that relates the
nitrogen-to-oxygen abundance ratio N/O in an H\,{\sc ii} region with
the intensities of the strong lines in its spectrum.  Panel $a$ of
Fig.~\ref{figure:nolin} shows the N/O ratio as a function of the line
intensity ratio N$_{2}$/R$_{2}$ for our sample of reference H\,{\sc
ii} regions.  The H\,{\sc ii} regions were subdivided into three
subsamples according to the intensity of the $N_{2}$ line.  Those
subsamples are plotted using different symbols.  Panel $a$ of
Fig.~\ref{figure:nolin} suggests that the relation between the
abundance ratio N/O and the line intensity ratio N$_{2}$/R$_{2}$ can
be fitted by a linear expression for each subsample of  H\,{\sc ii}
regions.  The N/O -- $N_{2}/R_{2}$ relations for different subsamples
are shifted relative each to other and have different slopes.  We 
derived the following calibration relation for the total sample: 
\begin{eqnarray}
       \begin{array}{lll}
\log {\rm (N/O)}  & = &  -0.657 - 0.201 \, \log N_{2}  \\  
                  & + & (0.742 -0.075 \, \log N_{2}) \times \log(N_{2}/R_{2})  \\ 
     \end{array}
\label{equation:nolin}
\end{eqnarray}
The lines in panel $a$ of Fig.~\ref{figure:nolin} show the
calibration relation for different values of the intensity of the 
$N_{2}$ line.

Panel $b$ of Fig.~\ref{figure:nolin} shows the nitrogen-to-oxygen
abundance ratio (N/O)$_{calibr}$ derived from the calibration relation
(Eq.~\ref{equation:nolin}) as a function of the nitrogen-to-oxygen
abundance ratio (N/O)$_{T_{e}}$ for the calibrating H\,{\sc ii}
regions.  The grey points mark individual H\,{\sc ii} regions.  We
again split our sample of H\,{\sc ii} regions into bins of 0.1 dex in
the (N/O)$_{T_{e}}$ abundance ratio.  We determined the mean
nitrogen-to-oxygen abundance ratios (N/O)$_{T_{e}}^{mean}$ and
(N/O)$_{calibr}^{mean}$ for the H\,{\sc ii} regions in each bin.  The
absolute values of the mean difference between (N/O)$_{calibr}$ and
(N/O)$_{T_{e}}$ were estimated for each bin using
Eq.~\ref{equation:meandif}.  Those mean values are shown in the panel
$b$ of Fig.~\ref{figure:nolin} by the dark points while the bars show
the absolute values of the mean difference between the
nitrogen-to-oxygen abundance ratios (N/O)$_{calibr}$ and
(N/O)$_{T_{e}}$ for each bin.  The solid line is that of equal values,
the dashed lines show the $\pm$0.1 dex deviations from the unity line.

Panel $c$ of Fig.~\ref{figure:nolin} shows the normalized histogram of
the differences between the N/O ratios derived from the calibration
relation (Eq.~\ref{equation:nolin}) and through the $T_{e}$ method for
our calibrating H\,{\sc ii} regions. 

Above, we constructed the calibration relations for the
determination of the nitrogen and oxygen abundances.  If the nitrogen
and oxygen abundances are estimated separately then the
nitrogen-to-oxygen abundance ratio can be easily obtained.  Let us
compare the nitrogen-to-oxygen abundance ratio (N/O)$_{calibr}$
determined immediately from the corresponding calibration relation
(Eq.~\ref{equation:nolin}) and the nitrogen-to-oxygen abundance ratio
(N/O)$_{R}$ obtained from the nitrogen (N/H)$_{R}$ and oxygen
(O/H)$_{R}$ abundances determined separately from the corresponding
relations (Eqs.~\ref{equation:ohru}, \ref{equation:ohrl} for the
oxygen abundances and Eqs. \ref{equation:nhru}, \ref{equation:nhrl}
for the nitrogen abundances).  Panel $d$ of Fig.~\ref{figure:nolin}
shows the comparison between the nitrogen-to-oxygen abundance ratio
(N/O)$_{calibr}$ and (N/O)$_{R}$.  

Inspection of Fig.~\ref{figure:nolin} suggests that the N/O abundance
ratio in an H\,{\sc ii} region estimated from the intensities of the
two strong emission lines $N_{2}$ and $R_{2}$ agrees with the directly
measured N/O abundance ratio within $\sim$0.05 dex for the bulk of our
reference H\,{\sc ii} regions.  The mean difference between the N/O
abundance ratios obtained from Eq.~\ref{equation:nolin} and through
the $T_{e}$ method is 0.026 dex for our 313 calibrating data points. 
The values of the nitrogen-to-oxygen abundance ratio (N/O)$_{calibr}$
determined directly from the corresponding calibration relation and
the nitrogen-to-oxygen abundance ratio (N/O)$_{R}$ obtained from 
nitrogen (N/H)$_{R}$ and oxygen (O/H)$_{R}$ abundances determined
separately are also close to each other.  This suggests that our
calibration relations are self-consistent.

\section{Discussion and conclusions}

In this study, we derived simple relations (calibrations) between the
oxygen (nitrogen) abundance in H\,{\sc ii} regions using the
intensities of the strong lines in their spectra.  These relations can
be applied to derive nebular abundances from spectroscopic
measurements that do not permit the use of the direct method for
abundance determinations.  The choice of more sophisticated
expressions may result in an even better agreement between the
calibration-based and the directly measured abundances. However, the
abundances produced by the suggested calibrations agree with the
measured abundances to within about 0.1 dex, which is comparable with
the expected range of uncertainties in the directly measured
abundances of the calibrating H\,{\sc ii} regions.  Therefore, one can
expect that a significant increase of the precision of the
calibration-based abundances can only be provided by a calibration
based on a sample of reference H\,{\sc ii} regions that is larger in
quantity and/or higher in the quality of the abundance determinations
through the direct $T_{e}$-based method as compared to the sample used
here.  For that purpose, new high precision measurements of H\,{\sc
ii} regions would be needed. 

In the following we discuss the properties and validity of the
suggested calibrations.

\subsection{The transition between the upper and lower branches}

\begin{figure}
\resizebox{1.00\hsize}{!}{\includegraphics[angle=000]{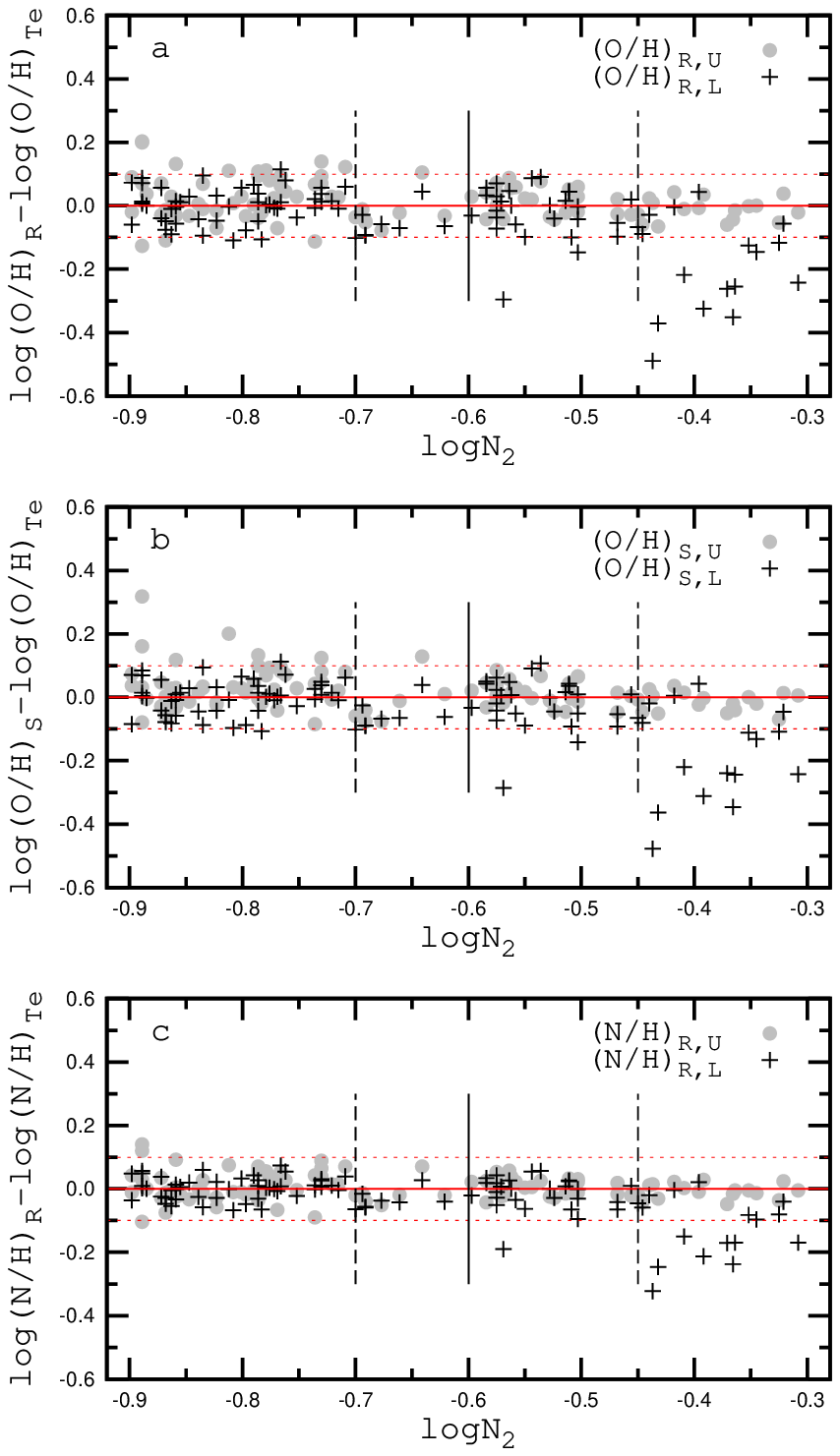}}
\caption{Comparison of the abundances in H\,{\sc ii} regions with
log$N_{2}$ near the boundary value obtained through the calibration
relations for the upper and lower branches.  Panel $a$. The
differences between oxygen abundances estimated from the $R$
calibration for the upper branch and the direct $T_{e}$-based oxygen 
abundance (O/H)$_{R,U}$ -- (O/H)$_{T_{e}}$ (circles) and the difference
(O/H)$_{R,L}$ -- (O/H)$_{T_{e}}$ (plus signs) as a function of the
nitrogen line $N_{2}$ intensity.  The solid vertical line marks the
adopted boundary dividing the ranges of the applicability of the lower
and upper branch calibration relations.  The dashed vertical lines show
the interval in log$N_{2}$ where both relations produce abundances
that are in close agreement.  Panel $b$. The same as panel $a$ 
but for the $S$ calibration.  Panel $c$. The same as panel $a$
but for nitrogen abundances. }
\label{figure:boundary}
\end{figure}

We constructed different calibration relations for the abundance
determinations in H\,{\sc ii} regions of the upper and lower branches.
A value of log$N_{2} = -0.6$ was adopted as the condition separating
the ranges of applicability of the calibration relations.  H\,{\sc ii}
regions with log$N_{2} \ge -0.6$ are assumed to belong to the upper
branch and objects with log$N_{2} < -0.6$ to the lower branch. This
separation criterium was inferred from the examination of
Fig.~\ref{figure:ln2xhte} and is somewhat arbitrary.  It is important
that the abundances produced by the calibrations for the lower and
upper branches are compatible with each other in the boundary region
separating the ranges of their applicability.  Indeed, two H\,{\sc ii}
regions located close to the dividing line on both sides have similar
abundances.  Therefore, the abundance resulting from the lower branch
calibration in an H\,{\sc ii} region located on the one side of the
boundary line should be close to the abundance produced by the upper
branch calibration in an H\,{\sc ii} region located on the other side 
of the dividing line.  Is this the case?

We now consider H\,{\sc ii} regions with nitrogen line $N_{2}$
intensities near the adopted boundary value, with intensities from
log$N_{2} = -0.9$ to $-0.3$. As a test, the abundances in
those H\,{\sc ii} regions were obtained both with the calibration
relations for the upper branch and for the lower branch. The filled
grey circles in panel $a$ of Fig.~\ref{figure:boundary} show the
differences between the oxygen abundance estimated from the $R$
calibration for the upper branch and the direct $T_{e}$-based oxygen
abundance (O/H)$_{R,U}$ -- (O/H)$_{T_{e}}$ as a function of the
nitrogen line $N_{2}$ intensity. The plus signs show the differences
between the oxygen abundance estimated from the $R$ calibration for
the lower branch and the direct $T_{e}$-based oxygen abundance
(O/H)$_{R,L}$ -- (O/H)$_{T_{e}}$ for the same objects. The solid
vertical line shows the adopted boundary dividing the ranges of the
applicability of the lower and upper branch calibration relations. The
dashed vertical lines indicate the interval in log$N_{2}$ where both
relations produce rather close abundances.  Panel $a$ of
Fig.~\ref{figure:boundary} shows that the $R$ calibration for the
upper branch provides rather accurate oxygen abundances for objects
with log$N_{2}$ below the adopted limit, up to log$N_{2} \sim -0.7$.
In turn, the $R$ calibration for the lower branch results in fairly
accurate oxygen abundances for objects with log$N_{2}$ higher than the
adopted limit, up to log$N_{2} \sim -0.45$, with a few exceptions. The
ranges of applicability of the $R$ calibration relations for the
oxygen abundance determinations overlap in the range from log$N_{2}
\sim -0.7$ to $-0.45$.  The dashed vertical lines show the interval in
log$N_{2}$ where both relations produce abundances that are in close
agreement.

Panel $b$ of Fig.~\ref{figure:boundary} shows the difference
(O/H)$_{S,U}$ -- (O/H)$_{T_{e}}$ and (O/H)$_{S,L}$ -- (O/H)$_{T_{e}}$
as a function of the  $N_{2}$ nitrogen line intensity. The ranges of
applicability of the $S$ calibration relations for the oxygen
abundance determinations also overlap.  Panel $c$ of
Fig.~\ref{figure:boundary} shows the difference (N/H)$_{R,U}$ --
(N/H)$_{T_{e}}$ and (N/H)$_{R,L}$ -- (N/H)$_{T_{e}}$ as a function of
the $N_{2}$ nitrogen line intensity. Again, the ranges of
applicability of the $R$ calibration relations for the nitrogen 
abundance determinations overlap.

Thus, the calibrations for the lower and upper branches are compatible
with each other in the boundary regime dividing the ranges of their
applicability.  Moreover, the ranges of their applicability overlap.
In practice, H\,{\sc ii} regions with $N_{2}$ nitrogen line intensities
near the adopted boundary value can be misclassified due to errors in
the $N_{2}$ measurements.  The H\,{\sc ii} regions that truly belong
to the upper branch with underestimated  $N_{2}$ nitrogen line
intensities can be classified as objects of the lower branch and vice
versa.  However, since the ranges of the applicability of the lower
and upper branch calibration relations overlap the error in the
abundance due to the misclassification of H\,{\sc ii} regions is not
large.

\subsection{Verification of the calibrations: abundances in a large 
sample of  H\,{\sc ii} regions }

\begin{figure*}
\resizebox{1.00\hsize}{!}{\includegraphics[angle=000]{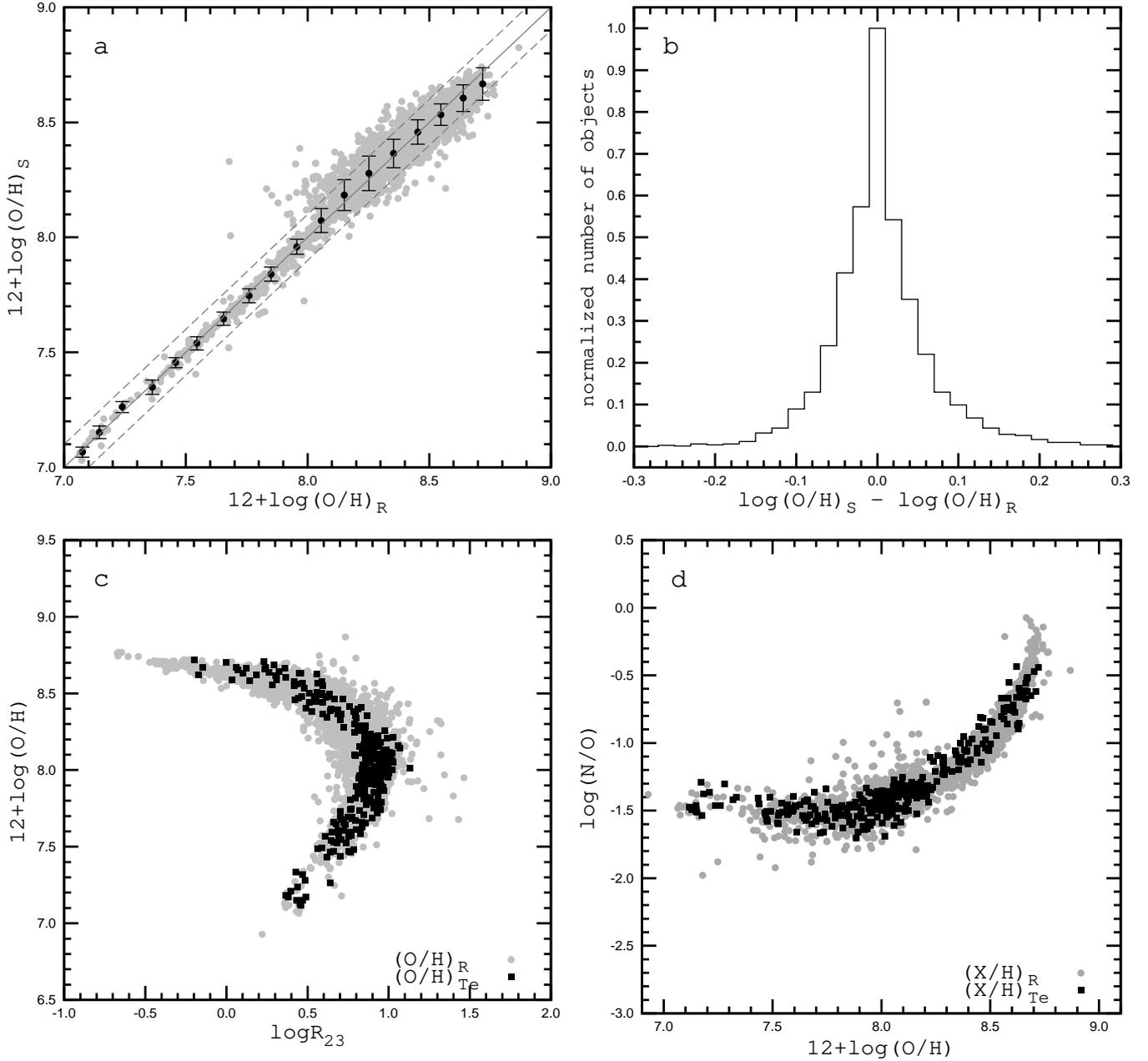}}
\caption{The application of the suggested calibrations to a large
sample of 3454 spectra of H\,{\sc ii} regions in spiral and irregular
galaxies.  Panel $a$.  Oxygen abundance (O/H)$_{S}$ vs.\ oxygen
abundance (O/H)$_{R}$. The grey points denote individual H\,{\sc ii}
regions.  The dark points represent mean abundances for objects in bins
with a size of 0.1 dex in (O/H)$_{R}$.  The error bars show the mean
value of the differences between (O/H)$_{S}$ and (O/H)$_{R}$.  The
solid diagonal line indicates equality; the dashed lines show the
$\pm$0.1 dex offsets from equal values.  Panel $b$.  The normalized
histogram of the differences between (O/H)$_{S}$ and (O/H)$_{R}$.
Panel $c$.  The O/H -- $R_{23}$ diagram. The grey points are the
(O/H)$_{R}$ abundances for the compiled sample of H\,{\sc ii} regions. 
The dark squares mark the (O/H)$_{T_{e}}$ abundances for
the calibrating data points.  Panel $d$. The N/O -- O/H diagram.
The grey points are the (X/H)$_{R}$ abundances for the compiled
sample.  The dark squares are the (X/H)$_{T_{e}}$ abundances
for the calibrating data points. 
}
\label{figure:r23oh}
\end{figure*}

We have compiled a large number of spectra of H\,{\sc ii} regions in
spiral and irregular galaxies in our previous studies
\citet{Pilyugin2012,Pilyugin2014}.  The $R_{2}$, $R_{3}$, $N_{2}$, and
$S_{2}$ line intensity measurements are available in 3454 spectra of
H\,{\sc ii} regions.  Those data provide an additional possibility to
test the validity of the abundances produced by the suggested
calibrations.

Panel $a$ of Fig.~\ref{figure:r23oh} shows the $S$-calibration-based
oxygen abundances (O/H)$_{S}$ as a function of $R$-calibration-based
oxygen abundances (O/H)$_{R}$ for the compiled sample of H\,{\sc ii}
regions.  The grey points stand for individual H\,{\sc ii} regions.
The dark points represent the mean abundances for objects in bins with
sizes of 0.1 dex in (O/H)$_{R}$.  The error bars show the mean value
of the differences between (O/H)$_{S}$ and (O/H)$_{R}$ in the H\,{\sc
ii} regions within each bin.  Objects with large absolute values of
the difference between (O/H)$_{S}$ and (O/H)$_{R}$ abundances (i.e.,
larger than 0.2 dex) were excluded in the determinations of the mean
values of the abundance and mean values of the abundance differences.
The solid diagonal line represents equality; the dashed lines show
$\pm$0.1 dex offsets from equal values.  Panel $b$ of
Fig.~\ref{figure:r23oh} displays the normalized histogram of the
differences between the (O/H)$_{S}$ and (O/H)$_{R}$ abundances.  The
panels $a$ and $b$ of Fig.~\ref{figure:r23oh} demonstrate that the
(O/H)$_{S}$ and the (O/H)$_{R}$ abundances agree with each other
within $\sim 0.05$ dex for the majority of the  H\,{\sc ii} regions.
It should be emphasized the value of 0.05 dex cannot be interpreted as
the precision of the abundance determinations with our calibration
relations.  The uncertainties in the $R_{3}$ and $N_{2}$ line
measurements can introduce similar errors in the (O/H)$_{R}$ and
(O/H)$_{S}$ abundances.  Therefore, this abundance uncertainty cannot
be revealed through the comparison between (O/H)$_{R}$ and (O/H)$_{S}$
abundances.  

The O/H vs.\ $R_{23}$ diagram is a well-studied diagnostic diagram.
This diagram was used to construct calibrations for oxygen abundance
determinations by numerous investigators
\citep[e.g.,][]{Edmunds1984,Dopita1986,McGaugh1991,Zaritsky1994,Pilyugin2000,Pilyugin2001a,Pilyugin2005}.
Panel $c$ of Fig.~\ref{figure:r23oh} shows the O/H -- $R_{23}$
diagram. The grey points indicate the (O/H)$_{R}$ abundances for our
sample of compiled H\,{\sc ii} regions.  The dark squares mark the
(O/H)$_{T_{e}}$ abundances for the calibrating data points.  The
objects with calibration-based (O/H)$_{R}$ abundances follow closely
the trend traced by objects with direct (O/H)$_{T_{e}}$ abundances. 

The N/O vs.\ O/H diagram can be also used to test the validity of the
oxygen and nitrogen abundances produced by the suggested calibrations.
Many studies are devoted to the investigation of the N/O -- O/H
diagram \citep[][among many
others]{Edmunds1978,Izotov1999,Henry2000,Pilyugin2003,Pilyugin2004,Berg2012,Annibali2015}.
Panel $d$ of Fig.~\ref{figure:r23oh} shows the N/O -- O/H diagram.
The grey points mark the (X/H)$_{R}$ abundances for our compiled
sample.  The dark squares show the (X/H)$_{T_{e}}$ abundances for the
calibrating data points.  Again, the objects with $R$-based abundances
follow closely the trend traced by objects with direct $T_{e}$-based
abundances.  It is commonly accepted that the break in the N/O -- O/H
diagram is caused by fact that since 12 +log(O/H) $\ga 8.2$, secondary
nitrogen becomes dominant and the nitrogen abundance increases at a
faster rate than the oxygen abundance \citep{Henry2000}.  The scatter
in the N/O -- O/H diagram can be caused by the time delay between
nitrogen and oxygen enrichment, the local enrichment in nitrogen by
Wolf-Rayet stars, and by enriched galactic winds
\citep[e.g.][]{Edmunds1978,Pilyugin1992,Pilyugin1993,Henry2000,LopezSanchez2010,PilyuginThuan2011,Tsujimoto2013,MirallesCaballero2014}.

\subsection{Verification of the calibrations: abundance gradients in 
the galaxy M~101}

\begin{figure}
\resizebox{1.00\hsize}{!}{\includegraphics[angle=000]{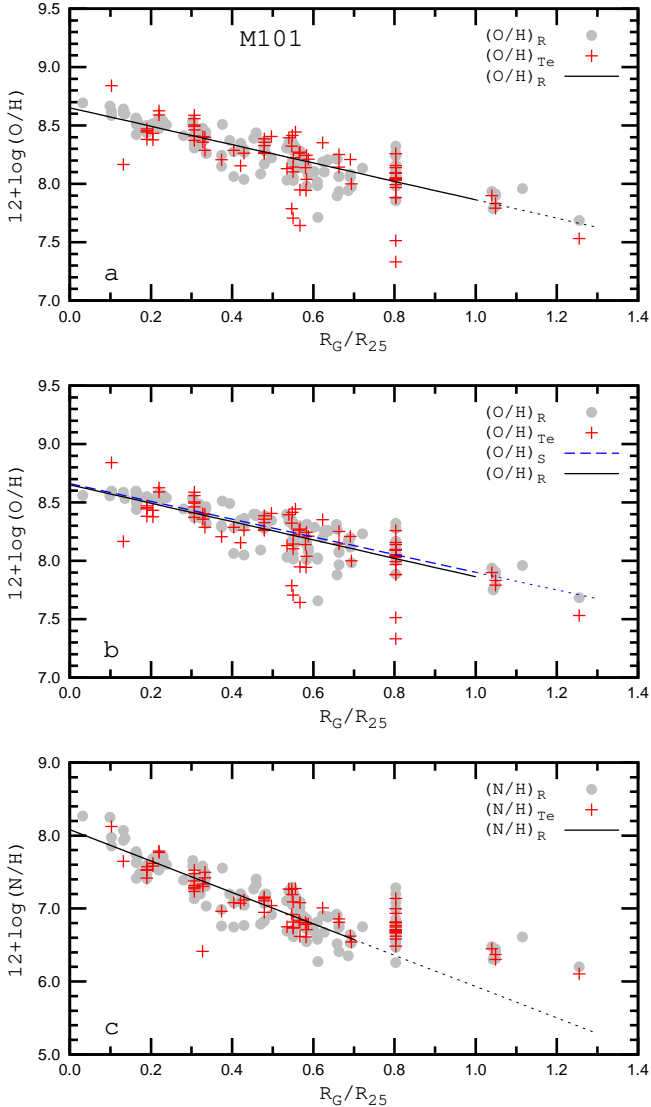}}
\caption{Panel $a$. The radial distributions of the oxygen
(O/H)$_{R}$ (circles) and (O/H)$_{T_{e}}$ abundances (plus signs) 
across the disk of the galaxy M~101.  The X-axis shows the
galactocentric radius $R_G$ along M~101's disk normalized by the the optical
$R_{25}$ radius.  The solid line indicates the linear best fit
(O/H)$_{R}$ = $f$($R_{G}$) to objects within the optical radius
$R_{25}$.  The dotted line is the extrapolation of the best fit beyond
the optical radius.  Panel $b$. The radial distributions of the
oxygen (O/H)$_{S}$ (circles) and (O/H)$_{T_{e}}$ (plus signs)
abundances.  The grey (blue) dashed line shows the linear best fit
(O/H)$_{S}$ = $f$($R_{G}$) to the objects within the optical radius
$R_{25}$, and the dotted line is its extrapolation beyond the optical
radius.  The solid line comes from panel $a$.  Panel $c$.  The
radial distributions of the nitrogen (N/H)$_{R}$ (circles) and
(N/H)$_{T_{e}}$ abundances (plus signs).  The solid line is the linear
best fit (N/H)$_{R}$ = $f$($R_{G}$) to the objects with galactocentric
distances less than $0.7R_{25}$, and the dotted line is its
extrapolation beyond this radius. 
}
\label{figure:m101}
\end{figure}

The oxygen abundance distribution across the disk of the galaxy M~101
($\equiv$ NGC~5457) is considered in a number of investigations
\citep[e.g.,][]{GarnettKennicutt1994,KennicuttGarnett1996,Pilyugin2001b,LiBresolin2013}.
The galaxy M~101 is an attractive object to test the validity of
different calibrations for two reasons. First, the auroral lines were
measured in the spectra of a number of its H\,{\sc ii} regions and,
consequently, the abundances in those H\,{\sc ii} regions could be
derived through the direct $T_{e}$ method.  This provides a possibility
to compare the abundances produced by empirical calibrations with
these direct abundances, i.e., in some sense the galaxy M~101 can be
considered as a ``Rosetta stone''.  Second, the oxygen abundances of
the H\,{\sc ii} regions of M~101 cover a large range of abundances
spanning approximately an order of magnitude.  This provides a
possibility to test the validity of our calibrations over the large
range of abundances.  Here we use emission-line measurements in 142
spectra of  H\,{\sc ii} regions in M~101 taken from
\citet{Hawley1978,SedwickAller1981,Rayo1982,Skillman1985,TorresPeimbert1989,KinkelRosa1994,
GarnettKennicutt1994,KennicuttGarnett1996,vanZee1998,Luridiana2002,Kennicutt2003,Bresolin2007,Esteban2009,LiBresolin2013}.
It should be noted that the number of spectra is larger than the
number of measured H\,{\sc ii} regions because some H\,{\sc ii}
regions were measured several times.  

Panel $a$ of Fig.~\ref{figure:m101} shows the radial distributions of
the $T_{e}$-based (plus signs) and $R$-calibration-based (circles)
oxygen abundances across the disk of the galaxy M~101.  The solid line
shows the linear best fit (O/H)$_{R}$ = $f$($R_{G}$) to the objects
within the optical radius $R_{25}$.  The dotted line is the
extrapolation of the best fit beyond the optical radius.  Panel $a$ of
Fig.~\ref{figure:m101} shows that the radial distribution of the
$R$-calibration-based oxygen abundances follows that of the
$T_{e}$-based abundances very well.  The radial distributions of the
oxygen abundances within and beyond the optical radius (up to $\sim
1.3 R_{25}$) can be described by the unique (O/H)$_{R}$ -- $R_{G}$
relation. 

Panel $b$ of Fig.~\ref{figure:m101} shows the radial distributions of
the oxygen (O/H)$_{S}$ (circles) and (O/H)$_{T_{e}}$ (plus signs)
abundances.  The grey (blue) dashed line is the linear best fit
(O/H)$_{S}$ = $f$($R_{G}$) to the objects within the optical radius
$R_{25}$, and the dotted line is its extrapolation beyond the optical
radius.  For comparison, the best fit to the (O/H)$_{R}$ abundances is
also shown by the solid line (from panel $a$).  The (O/H)$_{S}$ --
$R_{G}$ relation is very close to the (O/H)$_{R}$ -- $R_{G}$ relation;
in fact they coincide with each other. 

Panel $c$ of Fig.~\ref{figure:m101} shows the radial distributions of
the nitrogen (N/H)$_{R}$ (circles) and (N/H)$_{T_{e}}$ (plus signs)
abundances.  The solid line is the linear best fit (N/H)$_{R} =
f(R_{G}$) to the objects with galactocentric distances less than $0.7
R_{25}$, and the dotted line is its extrapolation beyond this radius.
As in the case of the oxygen abundances, the radial distribution of
the $R$-calibration-based nitrogen abundances follows those of the
$T_{e}$-based nitrogen abundances well.  In contrast to the oxygen
abundance, however, the radial distribution of the nitrogen abundances
shows a break at $\sim$0.7$R_{25}$ (it is difficult to derive the
exact value of the break radius because of the scatter in the nitrogen
abundance at a given galactocentric distance).  The slope of the radial
distribution of the nitrogen abundances beyond this radius becomes
shallower.  This is not surprising.  As we noted above, since 12
+log(O/H) $\ga 8.2$, secondary nitrogen becomes dominant and the
nitrogen abundance increases at a faster rate than the oxygen
abundance.

\subsection{On the metallicity scale for H\,{\sc ii} regions}

For the sake of clarity we will briefly discuss the validity and
reliability of the oxygen abundances. The validity and reliability of
the calibration-based abundances can be addressed on three levels.
Firstly, the accuracy of the calibration relations depend on the
number and quality of the calibrating data points.
Fig.~\ref{figure:gistxhe} shows that the number of the reference
H\,{\sc ii} regions at high metallicities is relatively small. As was
noted earlier, new high precision measurements of H\,{\sc ii} regions,
especially at high metallicities, would be desirable to increase the
reliability of the calibration-based abundances.  The abundances in
H\,{\sc ii} regions provided by the suggested calibration relations
are in agreement with the $T_{e}$-based abundances within 0.1 dex,
i.e., the relative accuracy of the calibration-based abundances are
within 0.1 dex.  There is no systematic discrepancy between
calibration-based and $T_{e}$-based abundances. Therefore, the
validity of the absolute calibration-based abundances (the potential
systematic error) is, in fact, determined by the validity of the
$T_{e}$-based metallicity scale for H\,{\sc ii} regions. 

Secondly, there is a discrepancy between the values of the abundance
of a given H\,{\sc ii} region derived through the $T_{e}$ method and
via the model fitting using the same collisionally excited lines (see
references in Section 1).  
The validity and reliability of abundances obtained both
through model fitting as well as through the $T_{e}$ method can be 
questioned. On the one hand, it is pertinent to quote a passage from a
review paper written by an H\,{\sc ii} region modeller
\citet{Stasinska2004}:  ``A widely spread opinion is that
photoionization model fitting provides the most accurate abundances.
This would be true if the constraints were sufficiently numerous (not
only on emission line ratios, but also on the stellar content and on
the nebular gas distribution) and if the model fit were perfect (with
a photoionization code treating correctly all the relevant physical
processes and using accurate atomic data). These conditions are never
met in practice.'' On the other hand, there are a number of factors
that can affect $T_{e}$-based abundance determinations.  It should be
stressed that the underlying logic of the classical $T_{e}$ method is
irreproachable.  However, the practical realization of the $T_{e}$
method relies on some assumptions that can be questioned.  Indeed, the
equations of the $T_{e}$ method correlate the intensities of the
emission lines with the electron temperature and abundances within a
volume with constant conditions (electron temperature and chemical
composition). In reality, possible spatial variations of the physical
conditions inside an H\,{\sc ii} region can affect the $T_{e}$-based
abundance determinations. 

Thirdly, usually collisionally excited lines are used for the
determination of the electron temperature and abundance in H\,{\sc ii}
regions, e.g., in the $T_{e}$ method.  The electron temperature of an
H\,{\sc ii} region can be also derived from the Balmer (or Paschen)
jump \citep{Peimbert1967}, whereas the abundance of an H\,{\sc ii}
region can also be determined from optical recombination lines.
\citet{Guseva2006} and \citet{Guseva2007} determined the Balmer and/or
Paschen jump temperatures in a large sample of low-metallicity (12 +
log(O/H) $\la 8.36$) H\,{\sc ii} regions.  They found that the
temperatures of the O$^{++}$ zones determined through the equation of
the $T_{e}$ method (from collisionally excited lines) do not differ,
in a statistical sense, from the temperatures of the H$^{+}$ zones
determined from the Balmer and Paschen jumps although small
temperature differences of the order of 3\%--5\% cannot be ruled out.
The O$^{++}$ abundances obtained from the optical recombination lines
are systematically higher (by a factor of 1.3 -- 3) than the ones
determined through the equation of the $T_{e}$ method from
collisionally excited lines \citep[][and references
therein]{GarciaRojas2007,Esteban2014}.   

Different hypotheses were used to explain the discrepancy between the
abundances determined from the collisionally excited lines and from
the optical recombination lines.  \citet{Peimbert1967} assumed that
the temperature field within the  H\,{\sc ii} region is not uniform
but that there are instead the small-scale spatial temperature
fluctuations inside an H\,{\sc ii} region. If they are important then
the (O/H)$_{T_{e}}$ abundance would be a lower limit and the
abundances determined from the optical recombination lines remain
unaltered.  \citet{Tsamis2005} suggest a dual-abundance model
incorporating small-scale chemical inhomogeneities in the form of
hydrogen-deficient inclusions.  Neither abundances determined from the
collisionally excited lines nor from the optical recombination lines
are reliable in this case.  \citet{Nicholls2012} and
\citet{Dopita2013} argue that the energy distribution of electrons in
H\,{\sc ii} regions does not follow a Maxwell distribution.  In this
case, abundances determined from optical recombination lines are more
reliable. 

Thus, at the present time there is no absolute scale for the
metallicities of H\,{\sc ii} regions.  Until the problem of the
discrepancy between the abundances determined in different ways is
resolved, doubts about the validity and reliability of the
$T_{e}$-based (and any other) metallicity scale will remain. If at
some point irrefutable proof will be established that the
$T_{e}$-based abundances should be adjusted then our calibration
relations should be also reconsidered. 

Finally, it should be noted that the $T_{e}$ method (and,
consequently, our calibrations) produces the gas-phase oxygen
abundance in H\,{\sc ii} regions.  Some fraction of oxygen may be
embedded in dust grains.  \citet{Peimbert2010} have concluded that the
depletion of the oxygen abundance in H\,{\sc ii} regions can be around
0.1 dex increasing from $\sim 0.08$ dex in low-metallicity  H\,{\sc
ii} regions to $\sim 0.12$ dex in high-metallicity  H\,{\sc ii}
regions.  When this effect is taken into account the total (gas +
dust) oxygen abundance in an H\,{\sc ii} region is higher by $\sim
0.1$ dex than the one produced by the $T_{e}$ method and our
calibrations.

\section {Summary}

We derive simple expressions relating the oxygen abundance in H\,{\sc
ii} regions with the intensities of the three strong lines $R_{2}$,
$R_{3}$, and $N_{2}$ ($R$ calibrations) or $S_{2}$, $R_{3}$, and
$N_{2}$ ($S$ calibration). We also present the $R$ calibration for
nitrogen abundance determinations.  We use a sample of 313 reference
H\,{\sc ii} regions with $T_{e}$-based oxygen abundances obtained
originally for the counterpart method ($C$ method) as a sample of
calibrating data points. 

H\,{\sc ii} regions are usually divided in two classes: the
high-metallicity objects (the ``upper branch'') and low-metallicity
objects (the ``lower branch'').  We adopt a value of log$N_{2} = -0.6$
as boundary between the upper and lower branches.  In other words,
H\,{\sc ii} regions with log$N_{2} \ge -0.6$ belong to the upper
branch and H\,{\sc ii} regions with  log$N_{2} < -0.6$ are lower
branch objects.  We derive different calibration relations for the
abundance determinations in H\,{\sc ii} regions on the lower and upper
branches.  The ranges of the applicability of the calibration
relations for the upper and lower branches overlap in the boundary
region within a range of $-0.7 \la$ log$N_{2} \la -0.45$. 

The oxygen and nitrogen abundances estimated through the calibrations
agree with the $T_{e}$-based abundances within $\sim 0.1$ dex over the
whole metallicity range of  H\,{\sc ii} regions, i.e., the
relative accuracy of the calibration-based abundances is 0.1 dex. The
oxygen and nitrogen abundances in the high-metallicity  H\,{\sc ii}
regions (upper branch) can be estimated using only the intensities of
the two strong lines $R_{2}$ and $N_{2}$ (or $S_{2}$ and $N_{2}$ for
oxygen). 

The validity of the suggested calibrations was tested using a
compilation of more than three thousand spectra of H\,{\sc ii}
regions.  The locations of the calibration-based abundances in those
H\,{\sc ii} regions in the $R_{23}$ -- O/H and the N/O -- O/H digrams
follow well the general trends traced by H\,{\sc ii} regions with
$T_{e}$-based abundances.  Also, the radial distributions of the
calibration-based oxygen and nitrogen abundances across the disk of
the well-studied galaxy M~101 follow those of the $T_{e}$-based
abundances. 

We emphasize that there are two important advantages of our new
calibrations in comparison to the existing ones. Firstly, in our
approach, the oxygen abundances (O/H)$_{R}$ and (O/H)$_{S}$ --
produced by two different calibrations (based on different sets of
strong emission lines) -- agree within $\sim 0.05$ dex for the
majority of the H\,{\sc ii} regions.  
It should be noted that the value of 0.05 dex cannot be interpreted as
the accuracy of the abundance determinations through our calibration
relations.  Uncertainties in the  $R_{3}$ and $N_{2}$ line
measurements can introduce similar errors in the (O/H)$_{R}$ and
(O/H)$_{S}$ abundances.  Therefore, such uncertainties in the
abundance cannot be revealed through the comparison between
(O/H)$_{R}$ and (O/H)$_{S}$ abundances.  Secondly, 
since the ranges applicability of the calibration relations for the 
upper and lower branches overlap the problem with the abundance
determinations in the ``transition'' zone does not appear.

\section*{Acknowledgements}

We are grateful to the referee, Prof.\ M.A.~Dopita, for his 
constructive comments. \\
L.S.P.\ and E.K.G.\ acknowledge support in the framework of
Sonderforschungsbereich SFB 881 on ``The Milky Way System'' (especially
subproject A5), which is funded by the German Research Foundation
(DFG). \\ 
L.S.P.\ thanks for the hospitality of the Astronomisches
Rechen-Institut at Heidelberg University  where part of this
investigation was carried out. \\
This work was partly funded by the subsidy allocated to Kazan Federal
University for the state assignment in the sphere of scientific
activities (L.S.P.).  \\

\end{document}